\documentclass{LMCS}

\usepackage[latin1]{inputenc}
\usepackage{times}
\usepackage{latexsym}
\usepackage{amsmath}
\usepackage{amssymb}

\newcommand{\NC}{\newcommand}
\NC{\RNC}{\renewcommand}
\NC{\nev}{\newenvironment}
\NC{\rnev}{\renewenvironment}

\RNC{\labelenumi}{(\arabic{enumi})}
\RNC{\labelitemi}{\text{--}}
\RNC{\phi}{\varphi}
\RNC{\epsilon}{\varepsilon}
\NC{\bigmid}{\;\big|\;}
\NC{\Bigmid}{\;\Big|\;}
\RNC{\max}{\textup{max}}
\RNC{\min}{\textup{min}}
\RNC{\log}{\textup{log}\;}

\NC{\bende}{\eqno\qed}
\newlength{\probwidth}
\NC{\prob}[3][9]{\begin{center}\normalfont\fbox{\begin{tabular}[t]{rp{#1cm}}\textit{Input:}&#2.\\\textit{Problem:}&#3.\end{tabular}}\end{center}}

\NC{\pprob}[4][9]{\begin{center}\normalfont\fbox{\begin{tabular}[t]{rp{#1cm}}\textit{Input:}&#2.\\\textit{Parameter:}&#3\\\textit{Problem:}&#4.\end{tabular}}\end{center}}

\NC{\nprob}[4][9]{\begin{center}\normalfont\fbox{\addtolength{\probwidth}{#1cm}\parbox{\probwidth}{\textsc{#2}\\\hspace*{1.5em}\begin{tabular}[t]{rp{#1cm}}\textit{Input:}&#3.\\\textit{Problem:}&#4.\end{tabular}}}\end{center}}

\NC{\npprob}[5][9]{\begin{center}\normalfont\fbox{\addtolength{\probwidth}{#1cm}\parbox{\probwidth}{\textsc{#2}\\\hspace*{1.5em}\begin{tabular}[t]{rp{#1cm}}\textit{Input:}&#3.\\\textit{Parameter:}&#4\\\textit{Problem:}&#5.\end{tabular}}}\end{center}}

\RNC{\mod}[1]{\,(\textup{mod }#1)}
\NC{\col}{\textup{col}}
\NC{\id}{\textup{id}}
\NC{\tw}{\textup{tw}}
\NC{\FO}{\textup{FO}}
\NC{\MSO}{\textup{MSO}}
\NC{\LL}{\textup{L}}
\NC{\KK}{\textup{K}}
\NC{\CC}{\textup{C}}
\NC{\QQ}{\textup{Q}}
\NC{\var}{\textup{var}}
\NC{\voc}{\textup{voc}}
\NC{\free}{\textup{free}}
\NC{\dom}{\textup{dom}}
\NC{\rng}{\textup{rng}}
\NC{\ar}{\textup{ar}}
\NC{\CQ}{\textup{CQ}}

\NC{\fpcl}[1]{\left[#1\right]_{\text{\upshape fp}}}
\NC{\pr}{\leq^{\text{\normalfont fp}}_m}
\NC{\FPT}{\textup{FPT}}
\NC{\fpt}{\textup{fpt}}
\NC{\W}[1]{\text{$\textup{W}[#1]$}}
\NC{\A}[1]{\text{$\textup{A}[#1]$}}
\NC{\PTIME}{\textup{PTIME}}
\NC{\NP}{\textup{NP}}
\NC{\PSPACE}{\textup{PSPACE}}

\RNC{\Pr}{\textup{Pr}}

\NC{\MC}{\textup{MC}}
\NC{\pMC}{\textup{p-MC}}
\NC{\XK}{\textup{XK}}
\NC{\ORD}{\textup{ORD}}
\NC{\DTC}{\textup{DTC}}
\NC{\TC}{\textup{TC}}
\NC{\LFP}{\textup{LFP}}
\NC{\PFP}{\textup{PFP}}
\NC{\BOOL}{\textup{BOOL}}
\NC{\se}{\subseteq}
\NC{\new}[1][{\bfseries !}]{\marginpar{\raggedright\footnotesize #1}}

\newcommand{\str}[1]{\ensuremath{\mathcal #1}}

\def\doi{1 (1:2) 2005}
\lmcsheading%
{\doi}
{36}
{}
{}
{Sep.~\phantom{0}1, 2004}
{Mar.~\phantom{0}7, 2005}
{}

\begin{document}
\title{Model-Checking Problems as a Basis for Parameterized Intractability}
\author[J.~Flum]{J\"org Flum\rsuper a}
\address{{\lsuper a}Abteilung f\"ur Mathematische Logik\\
         Albert-Ludwigs-Universit\"at\\ 
         Eckerstr.~1\\
         79104 Freiburg\\
         Germany}
\email{flum@uni-freiburg.de}
\author[M.~Grohe]{Martin Grohe\rsuper b}
\address{{\lsuper b}Institut f\"ur Informatik\\ 
         Humboldt-Universit\"at\\
         Unter den Linden 6\\
         10099 Berlin\\
         Germany}
\email{grohe@informatik.hu-berlin.de}


\begin{abstract}
Most parameterized complexity classes are defined in terms of a
parameterized version of the Boolean satisfiability problem (the
so-called \emph{weighted satisfiability problem}). For example, Downey and
Fellow's W-hierarchy is of this form. But there are also classes such as 
the A-hierarchy, that are more naturally characterised in terms of
\emph{model-checking problems} for certain fragments of first-order logic.

Downey, Fellows, and Regan~(1998) were the first to establish a
connection between the two formalisms by giving a characterisation of the
W-hierarchy in terms of first-order model-checking problems. We improve their result and then prove a similar correspondence between weighted
satisfiability and model-checking problems for the A-hierarchy and the
$\text{W}^*$-hierarchy. Thus we obtain very uniform characterisations of many
of the most important parameterized complexity classes in both formalisms.

Our results can be used to give new, simple proofs of some of the core results
of structural parameterized complexity theory.
\end{abstract}

\maketitle

\section{Introduction}
Parameterized complexity theory allows a refined complexity analysis of
problems whose input consists of several parts of different sizes.  Such an
analysis is particularly well-suited for a certain type of logic based
algorithmic problems such as model-checking problems in automated verification
or database query evaluation. In such problems one has to evaluate a formula
of some logic in a finite structure.  Typical examples are the evaluation of
formulas of linear time temporal logic (LTL) in finite Kripke structures or
formulas of first-order logic (FO; relational calculus in database
terminology) in finite relational structures. Throughout this paper we adopt
the term \emph{model-checking problems} from verification when referring to
problems of this general type. It has turned out that usually the complexity
of these problems is quite high; for example, for both  LTL and FO, it is
PSPACE-complete \cite{siscla84,var82}. This high complexity of model-checking
problems is usually caused by large and complicated formulas. However, in the
practical situations in which model-checking problems occur one usually has to
evaluate a small formula in a very large structure. In our examples from
verification and database theory this is obvious. So an exponential time
complexity may still be acceptable as long as the exponential term in the
running time only involves the size of the input formula and not the much
larger size of the input structure. Lichtenstein and Pnueli~\cite{licpnu85}
argue along these lines to support their LTL-model-checking algorithm with a
running time of \mbox{$2^{O(k)}\cdot n$}, where $k$ is the size of the input
formula and $n$ the size of the input structure. While this argument just
follows algorithmic common sense, parameterized complexity theory, or more
precisely the theory of parameterized intractability, comes into play if
one wants to argue that no algorithm with a comparable running time exists for
FO-model-checking. Indeed, no algorithm for FO-model-checking with a running
time better than the trivial $n^{O(k)}$ is known, but classical complexity
theory does not provide the tools to show that no better algorithm exists.

So far we have argued that parameterized complexity theory is useful
for analysing certain algorithmic problems from logic. But it turns
out that the same logical problems are also very useful to lay a
foundation for parameterized complexity theory, and this is what the
present paper is about. 

Before describing our results, let us briefly recall the basic notions of
parameterized complexity theory. Instances of a parameterized problem consist
of two parts, which we call \emph{input} and \emph{parameter}. The idea is
that in the instances occurring in practice the parameter can be expected to
be small, whereas the input may be very large. For example, an instance of a
\emph{parameterized model-checking problem} consists of a structure and a 
formula, and we take the formula to be the parameter. Let $n$ denote the size
of the input of a parameterized problem and $k$ the size of the parameter.  A
parameterized problem is \emph{fixed parameter-tractable} if it can be solved
in time $f(k)\cdot p(n)$ for an arbitrary computable function $f$ and a polynomial
$p$.  FPT denotes the class of all fixed-parameter tractable problems. Just as
the Boolean satisfiability problem can be seen as the most basic intractable
problem in the classical theory of NP-completeness, a natural parameterization
of the satisfiability problem serves as a basis for the theory of
parameterized intractability: The \emph{weighted satisfiability problem} for a
class of Boolean formulas asks whether a given formula has a satisfying
assignment in which precisely $k$ variables are set to \textsc{true}; here $k$
is treated as the parameter. Unfortunately, it turns out that the complexity
of the weighted satisfiability problem is much less robust than that of the
unweighted problem. For example, the weighted satisfiability problem for
formulas in conjunctive normal form does not seem to have the same complexity
as the weighted satisfiability problem for arbitrary formulas. So instead of
getting just one class of intractable problems, we get a whole family of
classes of intractable parameterized problems each having a complete weighted
satisfiability problem. The most basic of these classes form the so-called
\emph{W-hierarchy}.

Downey, Fellows, and Regan~\cite{dowfelreg98} gave an alternative
characterisation of the W-hierarchy, which resembles Fagin's \cite{fag74} and
Stockmeyer's \cite{sto74} characterisation of the class NP and the polynomial
hierarchy. They proved that for each level $\W{t}$ of the W-hierarchy there is
a family $\Sigma_{t,u}[\tau]$, for $u\ge 1$, of classes of first-order
formulas of a certain vocabulary $\tau$ such that the model-checking problem
for each $\Sigma_{t,u}[\tau]$ is in $\W{t}$, and conversely each problem in
$\W{t}$ can be reduced to the model-checking problem for $\Sigma_{t,u}[\tau]$
for some $u\ge 1$. In \cite{flugro01} we improved this characterisation by
showing that $u$ can be taken to be $1$ and $\tau$ any vocabulary, which is
not unary. In other words, we showed that model-checking for
$\Sigma_{t,1}[\tau]$ is $\W{t}$-complete for any vocabulary $\tau$ that is not
binary. This result is the starting point for our present investigation. We
further improve the result by showing that the vocabulary $\tau$ can be taken
to be part of the input and does not have to be fixed in advance. This gives
us a very robust characterisation of the W-hierarchy in terms of first-order
model checking problems. To underline the significance of this
characterisation, we show that some of the most important structural results
on the W-hierarchy, the previously known proofs of which are very complicated
(cf.~Part~II of Downey and Fellow's monograph \cite{dowfel99}), can be derived
as easy corollaries of our results. Moreover, we derive a strengthening of the
so-called monotone and antimonotone collapse.

The correspondence between weighted satisfiability problems and model-checking
problems for first-order logic can be extended beyond the W-hierarchy. We
establish such a correspondence for the $\textup{W}^*$-hierarchy (introduced in
\cite{dowfeltay96}) and the A-hierarchy (introduced in \cite{flugro01}). For
each of these hierarchies a characterisation either in terms of weighted
satisfiability problems or in terms of model-checking problems was known
before; and for each of them we provide the counterpart.

The $\textup{W}^*$-hierarchy is a small variation of the W-hierarchy. As the
classes of the W-hierarchy, the classes of the $\textup{W}^*$-hierarchy are
defined via the weighted satisfiability problem; we give a characterisation in
terms of model-checking problems of first-order logic. It is an open problem
whether the W-hierarchy and the $\textup{W}^*$-hierarchy coincide. Downey,
Fellows, and Taylor were able to prove that $\W{1}=\text{W}^*[1]$
\cite{dowfeltay96} and $\W{2}=\text{W}^*[2]$ \cite{dowfel98}. The latter
result has a highly non-trivial proof; here we are able to derive
$\W{1}=\text{W}^*[1]$ and $\W{2}=\text{W}^*[2]$ as simple corollaries of our
characterisation of the $\textup{W}^*$-hierarchy. This gives a very
transparent proof of these results that also clearly shows why it only works
for the first two levels.

The A-hierarchy, which may be viewed as the parameterized analogue of the
polynomial hierarchy, is defined in terms of the parameterized halting problem
for alternating Turing machines. In \cite{flugro01}, we gave a
characterisation of the hierarchy in terms of model-checking problems for
fragments of first-order logic; in this characterisation the levels of the
A-hierarchy correspond to levels of quantifier alternation in first-order
formulas. Here we give a propositional characterisation in terms of the
\emph{alternating weighted satisfiability problem} (which may be viewed as the
parameterized version of the satisfiability problem for quantified Boolean
formulas). The overall picture that evolves is that in parameterized
complexity theory we have two different sources of increasing complexity: the
alternation of propositional connectives (leading to the W-hierarchy) and
quantifier alternation (leading to the A-hierarchy). Thus we
actually obtain a 2-dimensional family of parameterized classes which we call
the \emph{A-matrix} (see Figure~\ref{fig:matrix} on
page~\pageref{fig:matrix}). Each class of this matrix has natural
characterisations in terms of an alternating weighted satisfiability problem
and a model-checking problem for a fragment of first-order logic.  Let us
remark that in classical complexity, only quantifier alternation is relevant,
because the classes are closed under Boolean connectives. Thus there is only
the (1-dimensional) polynomial hierarchy.

In a last section, we use certain normal forms established here and a
known characterisation of the AW-hierarchy (introduced in \cite{abrdowfel95})
by first-order model-checking to give a simple proof of the collapse of the
AW-hierarchy to its first-level \cite{abrdowfel95}. Actually, we slightly
strengthen the result of \cite{abrdowfel95}. An application of this stronger
result can be found in \cite{frigro03}.

On a more technical level, our main contribution is a new and greatly
simplified proof technique for establishing the correspondence between
weighted satisfiability problems and model-checking problems. This technique
enables us to obtain all our results in a fairly uniform way. A major problem
in structural parameterized complexity theory is the lacking robustness of
most classes of intractable parameterized problems, leading to the abundance
of classes and hierarchies of classes. Maybe the technically most difficult
result of this paper is a normalisation lemma for the
relevant fragments of first-order logic which shows that the
vocabulary can be treated as part of the input of a model-checking
problem. 

\subsection*{Acknowledgements}
We are grateful to Catherine McCartin, Rod Downey, and Mike Fellows for
various discussions with both authors on the characterisation of the
A-hierarchy by alternating weighted satisfiability problems. These discussions
and our desire to understand the $\textup{W}^*$-hierarchy motivated us to
start the research that led to this paper. 

We would like to acknowledge that McCartin,
Downey, and Fellows already conjectured the characterisation of the A-hierachy
by alternating weighted satisfiability problems that we prove here.

\section{Preliminaries}\label{sec:pre}
In this section we recall some definitions and fix our notations.

 \subsection{Fixed-Parameter Tractability}\label{sub:pct}
A \emph{parameterized problem} is a set
$Q\subseteq\Sigma^*\times \Pi^*$, where $\Sigma$ and $\Pi$ are  finite
alphabets. If $(x,y)\in\Sigma^*\times\Pi^*$ is an instance of a
parameterized problem, we refer to $x$ as the \emph{input} and to $y$
as the \emph{parameter}. 

To illustrate our notation, let us give one example of a parameterized
problem, the \emph{parameterized clique problem} $p$-\textsc{Clique}:

\npprob{$p$-Clique}{A graph $\mathcal G$}{$k\in\mathbb N$ (say, in
  binary).}{Decide if $\mathcal G$ has a clique of size $k$}

\begin{defi}
A parameterized problem $Q\subseteq\Sigma^*\times\Pi^*$ is
\emph{fixed-parameter trac\-ta\-ble}, if there is a computable function
$f:\mathbb N\rightarrow\mathbb N$, a polynomial $p$, and an algorithm
that, given a pair $(x,y)\in\Sigma^*\times\Pi^*$, decides if
$(x,y)\in Q$ in at most $f(|y|)\cdot p(|x|)$ steps.

\FPT\ denotes the complexity class consisting of all fixed-parameter
tractable parameterized problems.  
\end{defi}

Occasionally we use the term \emph{fpt-algorithm} to refer to an algorithm
that takes as input pairs $(x,y)\in\Sigma^*\times\Pi^*$ and has a running time
bounded by $f(|y|)\cdot p(|x|)$ for some computable function
$f:\mathbb N\rightarrow\mathbb N$ and polynomial $p$. Thus a parameterized
problem is in FPT if it can be decided by an fpt-algorithm. However, we use
the term fpt-algorithm mostly when referring to algorithms computing mappings.

Complementing the notion of fixed-parameter tractability, there is a theory of
parameterized intractability. It is based on the following notion of
parameterized reduction:

\begin{defi}
An \emph{\fpt-reduction}  from the parameterized problem
$Q\subseteq\Sigma^*\times\Pi^*$ to the parameterized problem
$Q'\subseteq (\Sigma')^*\times(\Pi')^*$ is a mapping
$R:\Sigma^*\times\Pi^*\to(\Sigma')^*\times (\Pi')^*$ such that:
\begin{enumerate}
\item
For all
$(x,y)\in\Sigma^*\times\Pi^*$:
$
(x,y)\in Q\iff R(x,y)\in Q'.
$
\item
There is a computable function $g:\mathbb N\to\mathbb N$ such that for all
$(x,y)\in\Sigma^*\times\Pi^*$, say with $R(x,y)=(x',y')$, we have
$|y'|\le g(|y|)$. 
\item
$R$ can be computed by an fpt-algorithm.
\end{enumerate}
\end{defi}
We write $Q\le^{\fpt}Q'$ or simply $Q\le Q'$, if there is an
\fpt-reduction  from $Q$ to
$Q'$ and   set
\[
[Q]^{\fpt}:=\{Q'\mid Q'\le^{\fpt}Q\}.
\]
For a class C of parameterized problems, we let
    $$
[\textup{C}]^{\fpt}:=\bigcup_{Q\in \textup{C}}[Q]^{\fpt}.
$$

\subsection{Relational Structures and First-order Logic}\label{sub:fol}
A \emph{(relational) vocabulary} $\tau$ is a finite set of relation symbols. Each relation
symbol has an \emph{arity}. The {\em arity} of $\tau$ is the maximum of the
arities of the symbols in $\tau$.  A \emph{structure} $\mathcal
A$ of vocabulary $\tau$, or \emph{$\tau$-structure} (or, simply structure),
consists of a set $A$ called the {\em universe}, and an interpretation
$R^{\mathcal A}\subseteq A^r$ of each $r$-ary relation symbol $R\in\tau$. We
synonymously write $\bar a\in R^{\mathcal A}$ or $R^{\mathcal A}\bar a$ to
denote that the tuple $\bar a\in A^r$ belongs to the relation $R^{\mathcal
  A}$. For example, we view a {\em directed graph} as a structure $\mathcal
G=(G,E^{\mathcal G})$, whose vocabulary consists of one binary relation symbol
$E$. $\mathcal G$ is an (undirected) {\em graph}, if $E^{\mathcal G}$ is
irreflexive and symmetric. We
define the \emph{size} of a $\tau$-structure $\str A$ to be the number
\[
\|\str{A}\|:=|A|+\sum_{R\in\tau}\text{arity}(R)\cdot(|R^{\str A}|+1).
\]
$\|\str{A}\|$ is the size of a reasonable encoding of $\str{A}$ (see
\cite{flufrigro02} for details). For example, the size of a graph with $n$
vertices and $m$ edges is $O(n+m)$.

The class of all first-order formulas is denoted by $\FO$. They are built up
from atomic formulas using the usual boolean connectives and existential and
universal quantification.  Recall that \emph{atomic formulas} are formulas of
the form $x=y$ or $Rx_1\ldots x_r$, where $x,y,x_1,\ldots,x_r$ are variables
and $R$ is an $r$-ary relation symbol.  For $t\ge 1$, let $\Sigma_t$ denote the
class of all \FO-formulas of the form
\[
\exists x_{11}\ldots\exists x_{1k_1}\forall x_{21}\ldots\forall
x_{2k_2}\;\ldots\; Qx_{t1}\ldots Qx_{tk_t}\;\psi,
\]
where $Q=\forall$ if $t$ is even and $Q=\exists$ otherwise, and where $\psi$ is
quantifier-free. $\Pi_t$-formulas are defined analogously starting with
a block of universal quantifiers. Let $t, u \ge 1$. A formula
$\phi$ is $\Sigma_{t,u}$, if it is $\Sigma_{t}$ and all quantifier
blocks after the leading existential block have length $\le u$.
For example, a formula
\[
\exists x_1\ldots\exists x_k\forall y\exists z_1\exists z_2\psi,
\]
where $\psi$ is quantifier-free, is in $\Sigma_{3,2}$ (for every $k\ge 1$).

If $\mathcal A$ is a structure, $a_1,\ldots,a_n$ are elements of the universe
$A$ of $\mathcal A$, and $\phi(x_1,\ldots,x_n)$ is a first-order formula whose
free variables are among $x_1,\ldots,x_n$, then we write $\mathcal
A\models\phi(a_1,\ldots,a_n)$ to denote that $\mathcal A$ satisfies $\phi$ if
the variables $x_1,\ldots,x_n$ are interpreted by $a_1,\ldots,a_n$,
respectively. 

If $\Phi$ is a class of first-order formulas, then $\Phi[\tau]$ denotes the
class of all formulas of vocabulary $\tau$ in $\Phi$ and $\Phi[r]$, for $r\in
\mathbb N$, the class of all formulas in $\Phi$ whose vocabulary has arity
$\le r$.

If again $\Phi$ is a class of first-order formulas, then $p\textup{-MC}(\Phi)$
denotes the {\em (parameterized) model-checking problem for formulas in $\Phi$
}, i.e., the parameterized problem 
\npprob{$p\textup{-MC}(\Phi)$}{A structure
  \str{A}}{A sentence $\phi$ in $\Phi$}{Decide if \str{A} satisfies
  $\phi$}

Often, the natural formulation of a parameterized problem in first-order logic
immediately gives an \fpt-reduction to a model-checking problem, e.g.,
\begin{itemize}
\item $p\textsc{-Clique}\le p\textup{-MC}(\Sigma_{1}[2])$, since the existence of a clique of size $k$ is expressed by the $\Sigma_1$-sentence
\[
\exists x_1\ldots \exists x_k \bigwedge_{1\le i<j\le k}Ex_ix_j.
\]
\item $p\textsc{-Dominating Set}\le
  p\textup{-MC}(\Sigma_{2,1}[2])$. Here, $p\textsc{-Dominating Set}$ is the problem that asks if a graph \str{G} (the input) has a
  dominating set of size $k$ (the parameter); so we want to know if \str{G}
  satisfies the $\Sigma_{2,1}$-sentence
  \[
  \exists x_1\ldots \exists x_k\forall y
  (\bigwedge_{1\le i<j\le k} \neg x_i=x_j
  \wedge \bigvee_{1\le i\le k}(y=x_i\vee Eyx_i)).
  \]
\end{itemize}

\subsection{Propositional logic}
Formulas of propositional logic are important ingredients in the definitions
of various complexity classes of intractable parameterized problems. We recall
a few notions and fix our notations: Formulas of propositional logic are built
up from \emph{propositional variables} $X_1,X_2,\ldots$ by taking
conjunctions, disjunctions, and negations. The negation of a formula $\alpha$
is denoted by $\neg\alpha$. We distinguish between \emph{small conjunctions},
denoted by $\wedge$, which are just conjunctions of two formulas, and
\emph{big conjunctions}, denoted by $\bigwedge$, which are conjunctions of
arbitrary finite sets of formulas. Analogously, we distinguish between
\emph{small disjunctions}, denoted by $\vee$, and \emph{big disjunctions},
denoted by $\bigvee$.  A formula is \emph{small} if it neither contains big
conjunctions nor big disjunctions. By $\alpha=\alpha(Z_1,\ldots, Z_m)$ we indicate that  the variables in $\alpha$ are among $Z_1,\ldots, Z_m$.

Let $V$ be a set of propositional variables. We identify each assignment
\[S:V\to\{\textsc{true},\textsc{false}\}\] with the set $ \{X_i\in V\mid
S(X_i)=\textsc{true}\}\in 2^V.  $ The \emph{weight} of an assignment $S\in
2^V$ is $|S|$, the number of variables set to \textsc{true}.  A propositional
formula $\alpha$ is {\em $k$-satisfiable} (where $k\in\mathbb N$), if there is
an assignment for the set of variables of $\alpha$ of weight $k$
satisfying $\alpha$.

For a set $\Gamma$ of propositional formulas, the {\em weighted satisfiability
  problem \textsc{WSat}($\Gamma$) for formulas in $\Gamma$} is the following
parameterized problem: \npprob{\textsc{WSat}($\Gamma$)}{A propositional
  formula $\alpha\in\Gamma$}{$k\in\mathbb N$}{Decide if $\alpha$ is
  $k$-satisfiable}

The \emph{depth} of a formula is the maximum number of nested (big and small)
conjunctions and disjunctions appearing in this formula. The \emph{weft} of a
formula is the maximum number of nested big conjunctions and big disjunctions
appearing in it. Hence, the weft of a formula always is less  than or equal to its
depth. For $t,d\in \mathbb N$ with $t\le d$, we set
$$
\Omega_{t,d}:=\{\alpha \mid \mbox{propositional formula $\alpha$ has weft $\le t$ and depth $\le
  d$} \}.
$$
For $t\ge 0$ and $d\ge 1 $ define the sets $\Gamma_{t,d}$ and $\Delta_{t,d}$ by induction on $t$ (here, by  $(\lambda_1\land\ldots \land\lambda_r)$ we mean  the iterated small conjunction $((\ldots (\lambda_1\land\lambda_2)\ldots )\land\lambda_r)$):
\[
\begin{array}{rcl}
\Gamma_{0,d}&:=&\{(\lambda_1\land\ldots \land\lambda_r) \mid \lambda_1,\ldots ,\lambda_r \mbox{\ literals and $r\leq d$}\},\\
\Delta_{0,d}&:=&\{(\lambda_1\lor\ldots \lor\lambda_r) \mid \lambda_1,\ldots ,\lambda_r \mbox{\ literals and $r\leq d$}\},\\
\Gamma_{t+1,d}&:=&\{\bigwedge \Pi \mid \Pi\se \Delta_{t,d}\},\\
\Delta_{t+1,d}&:=&\{\bigvee \Pi \mid \Pi\se \Gamma_{t,d}\}.
\end{array}
\] 
If in the definition of $\Gamma_{0,d}$ and $\Delta_{0,d}$ we require that all
literals are positive (negative) we obtain the sets denoted by $\Gamma^+_{t,d}$
and $\Delta^+_{t,d}$ ($\Gamma^-_{t,d}$ and $\Delta^-_{t,d}$), respectively.
Clearly, $\Gamma_{t,d}\se\Omega_{t,t+d}$ and $\Delta_{t,d}\se\Omega_{t,t+d}$.
 
\section{Normalisation}\label{sec:nl}
We have introduced two logically defined families of parameterized problems,
the first based on model-checking problems for classes of first-order
sentences and the second based on the weighted satisfiability problem for
classes of propositional formulas. The main results of this paper establish a
tight correspondence between the two approaches; in fact, we present
formalisms that allow to translate from one family of parameterized problems
into the other. To prove these results, it is convenient to first simplify
each of the two sides separately.

\subsection{Propositional Normalisation}
The following lemma has been used by Downey, Fellows, and others as the first
step in numerous fpt-reductions (cf.~\cite{dowfel99}).

\begin{lem}[Propositional Normalisation]\label{lem:pn1} 
  Let $d\ge t\ge 0$. Then there is a polynomial time algorithm that computes
  for every formula in $\Omega_{t,d}$ an equivalent formula in
  $\Delta_{t+1,2^d}$.
\end{lem}

\medskip\noindent \textit{Proof (sketch): }We can restrict our attention to
formulas in $\Omega_{t,d}$ with negation symbols only in front of atomic
formulas. We proceed by induction on $t$: If $\alpha\in \Omega_{0,d}$, then $\alpha $ contains at most $2^d$ variables and we just compute an equivalent formula in disjunctive normal form. For $t\ge 1$, we  use the distributive law:
\begin{eqnarray*}
(\bigvee_{i\in I}\alpha_i \wedge \bigvee_{j\in J}\beta_j)&\mbox{is equivalent to}&\bigvee_{(i,j)\in I\times J}(\alpha_i\wedge \beta_j).
\end{eqnarray*}
\qed

\noindent
Note that the algorithm in Lemma~\ref{lem:pn1} is polynomial, because the
depth of the formulas is bounded by a fixed constant $d$. Obviously, no such
normalisation is possible for formulas of arbitrary depth. Even if the depth
of the formula is treated as a parameter, the reduction is not fixed parameter
tractable: the formula $\alpha'\in \Delta_{t+1,2^d}$ equivalent to a formula
$\alpha\in \Omega_{t,d}$ may have size $\Omega(|\alpha|^d)$. However,
as we shall see in Section~\ref{ss:star}, if we treat the depth as parameter
we can at least prove a weaker normalisation lemma (Lemma~\ref{lem:pnstar}).

\begin{cor}\label{cor:pn1} For all $d\ge t\ge 0$,
\[
\textsc{WSat}(\Omega_{t,d})\le\textsc{WSat}(\Delta_{t+1,2^d}).
\]
\end{cor}

\begin{rem}
Instead of propositional \emph{formulas}, Downey and Fellows
always work with Boolean \emph{circuits} (cf.~\cite{dowfel99}). However, since we are
only dealing with circuits and formulas of bounded depth, this does not really
make a difference. We can always transform circuits into
formulas in the most straightforward way. More precisely, if we define depth and weft of a
circuit in the natural way and denote by $C_{t,d}$ the class of all circuits
of weft $t$ and depth $d$, then we get the following results:
\begin{enumerate}
\item
Let $d\ge t\ge 0$. Then there is a polynomial time algorithm that
computes for every circuit in $C_{t,d}$ an equivalent formula
in $\Omega_{t,d}$.
\item Let $t\ge 0$. Then there is an fpt-algorithm that computes for every
  circuit in $C_{t,k}$ an equivalent formula in $\Omega_{t,k}$. Here $k$ is
  treated as the parameter.
\end{enumerate}
\end{rem}

\subsection{First-order normalisation}
The normalisation results for first-order logic presented in this subsection
are concerned with the vocabulary of the formulas in parameterized
model-checking problems. Actually, we prove that it is irrelevant, whether  we
consider arbitrary formulas or we restrict ourselves to a fixed vocabulary, as
long as it contains at least one binary relation symbol. This may not sound
very surprising, but is not easy to prove and was left open in our earlier
paper \cite{flugro01}.

The main results of this section are summarised in the following First-Order
Normalisation Lemma. To state the lemma we need two  more definitions: For all
$t,u\ge 1$, we call a $\Sigma_{t,u}$-formula
\[
\exists x_1\ldots \exists x_k \forall \bar y_1\ldots Q_t\bar y_t\phi
\] 
\emph{strict} if no atomic subformula of $\phi$ contains more than  one of the
variables $x_1,\ldots,x_k$. We denote the class of all strict
$\Sigma_{t,u}$-formulas by $\textup{strict-}\Sigma_{t,u}$. A $\Sigma_{t}$-formula is {\em simple}, if its quantifier-free part is a conjunction of literals 
in case $t$ is odd, and is a disjunction of literals 
in case $t$ is even.\footnote{Simple $\Sigma_{1}$-formulas are also called conjunctive queries with negation.}  We denote the class of all simple
$\Sigma_{t}$-formulas by $\textup{simple-}\Sigma_{t}$.

\begin{lem}[First-Order Normalisation Lemma]\label{lem:mcr}\mbox{}\\[-3ex]
\begin{enumerate}
\item 
For   $t\ge 2,u\geq 1$, 
$p\textup{-MC}(\Sigma_{t,u})\leq p\textup{-MC}(\textup{strict-}\Sigma_{t,1}[2]).$
\item
For all $t\ge 1$, $p\textup{-MC}(\Sigma_{t})\leq p\textup{-MC}(\textup{simple-}\Sigma_{t}[2]).$
\item 
$p\textup{-MC(FO)}\leq p\textup{-MC(FO}[2])$   (\cite{flugro01}).
\end{enumerate}
\end{lem}
The First-Order Normalisation Lemma is the only result of this section used in the rest of
the paper. Hence the reader not interested in its proof may pass to
Section~\ref{sec:bac} directly.

It will be useful to first recall the proof of (3) (from \cite{flugro01}) . We
then point out the difficulties in proving (2) by the same simple technique
and resolve these difficulties by Lemmas~\ref{lem:bin}--\ref{lem:sim}. The
proof of (1) is also complicated and will be carried out in several steps in
Lemmas~\ref{lem:arb}--\ref{lem:qua}.

\medskip\noindent \textit{Proof of Lemma~\ref{lem:mcr}(3):} Let
$(\str{A},\phi)$ be an instance of $p\textup{-MC}(\FO)$. We construct a
structure $\str{A}_b$ and a sentence $\phi_b\in\FO[2]$ such that
$(\str{A}\models \phi\iff \str{A}_b\models \phi_b) $.

Let $\tau$ be the vocabulary of \str{A}. We let $\str A_b$ be the {\em
  bipartite structure} or {\em incidence structure} associated with $\str A$:
Let $\tau_b$ be the vocabulary of arity 2 that contains a unary relation
symbol $P_R$ for every $R\in\tau$ and binary relation symbols
$E_1,\ldots,E_s$, where $s$ is the arity of $\tau$.  The universe $A_b$ of
$\str A_b$ consists of $A$ together with a new vertex $b_{R,\bar a}$ for all
$R\in\tau$ and $\bar a\in R^{\str A}$. The relation $E_i^{\str A_b}$ holds for
all pairs $(a_i,b_{R,a_1\ldots a_r})$, and $P_R^{\str A_b}:=\{b_{R,\bar
  a}\mid\bar a\in R^{\str A}\}$. Let $\phi_b$ be the $\FO$-sentence equivalent
to the $\tau'$-formula obtained from $\phi $ by replacing every atomic formula
$Rx_1\ldots x_r$ by (the simple $\Sigma_1$-formula)
\begin{equation}\label{pos}
\exists y(P_Ry \wedge E_1x_1y\wedge\ldots \wedge E_rx_ry).
\end{equation} 
Then clearly $(\str{A}\models \phi\iff \str{A}_b\models \phi_b) $. To see
that this construction yields an fpt-reduction, note that $\|\mathcal
A_b\|=O(\|\mathcal A\|)$. 
\qed

Why does the same construction not also work to get
$p\textup{-MC}(\Sigma_{t})\leq p\textup{-MC}(\Sigma_{t}[2])$?
Because if, say, a $\Sigma_1$ formula contains a negated atom $\neg Rx_1\ldots x_r$, then it will be replaced by a formula equivalent to
\begin{equation}\label{neg}
\forall y(\neg P_Ry \vee \neg E_1x_1y\vee \ldots \vee \neg E_rx_ry)
\end{equation}
and we obtain a formula that is no longer
equivalent to a $\Sigma_1$-formula. At first sight it seems that we can easily
resolve this problem by just extending the bipartite structure $\str A_b$ by
additional points $b_{\neg R,\bar a}$ for all $\bar a\not\in R^{\str A}$ and relation  symbols $P_{\neg R}$.
Unfortunately, in general the size of the resulting structure is not
polynomially bounded in the size of $\mathcal A$, since the vocabulary is not
fixed in advance.

We denote by
$\Sigma_{t}^+$ the class of all $\Sigma_{t}$-formulas without negation symbols and
by $\Sigma_{t}^-$ the class of all $\Sigma_{t}$-formulas in which there is a
negation symbol in front of every atom and there are no other negation symbols. Using the transition $(\str{A},\varphi)\mapsto(\str{A}_b,\varphi_b) $ we derive part (1) and (2) of the following lemma:

\begin{lem}\label{lem:bin}
\begin{enumerate}
\item
If $t\ge 1$ is odd, then 
\begin{eqnarray*}
p\textup{-MC}(\Sigma^+_{t})\leq p\textup{-MC}(\Sigma^+_{t}[2])&\mbox{and} & p\textup{-MC}(\textup{simple-}\Sigma^+_{t})\leq p\textup{-MC}(\textup{simple-}\Sigma^+_{t}[2]).
\end{eqnarray*}
\item
If $t\ge 1$ is even, then
\begin{eqnarray*}
p\textup{-MC}(\Sigma^-_{t})\leq p\textup{-MC}(\Sigma^-_{t}[2])&\mbox{and}& p\textup{-MC}(\textup{simple-}\Sigma^-_{t})\leq p\textup{-MC}(\textup{simple-}\Sigma^-_{t}[2]). 
\end{eqnarray*}
\item If $t\ge 1$ and $r\ge 1$, then $p\textup{-MC}(\textup{simple-}\Sigma_{t}[r])\leq p\textup{-MC}(\textup{simple-}\Sigma_{t}[2]).$
\end{enumerate}
\end{lem}
\proof
If $t$ is odd and $\varphi\in \Sigma^+_{t}$ (is simple), then the last quantifier block in $\varphi$ is existential (and the quantifier-free part is a conjunction of literals). Since $\varphi$ only has positive literals, in $\varphi_b$ this last existential block can absorb the quantifiers introduced by \eqref{pos} (and only further conjunctions are added to the quantifier-free part). Similarly, if $t$ is even and $\varphi\in \Sigma^-_t$ (is simple) , then the last quantifier block in $\varphi$ is universal (and the quantifier-free part is a disjunction of literals) and in $\varphi_b$ this  block can absorb the quantifiers introduced by \eqref{neg} (and only further disjunctions are added to the quantifier-free part).

It remains to prove (3). Fix $r\ge 1$. Given any structure \str{A} in a vocabulary $\tau$ of arity $r$, we obtain the structure $\str{A}'$ by adding  the complement of the relations  of \str{A}, more precisely: We  set $\tau':=\tau\cup\{R^c \mid R\in \tau \}\cup\{\not=  \}$ and we obtain $\str{A}'$ from $\str{A}$ setting  $(R^c)^{\str{A}'}:=A^{\textup{arity}(R)}\setminus R^{\str{A}}$ and $\not=^{\str{A}'}:=\{(a,b) \mid a,b\in A,\, a\not=b \} $.
Thus, $\|\str{A}'\|=O(\|\str{A}\|^r)$. The transition from $\str{A}$ to  $\str{A}'$ allows to replace in any formula   positive by negative literals and vice versa, thus showing that
\[
 p\textup{-MC}(\Sigma^+_{t}[r])\equiv^{\fpt}p\textup{-MC}(\Sigma_{t}[r]) \equiv^{\fpt} p\textup{-MC}(\Sigma^-_{t}[r])
\]
and
\[
 p\textup{-MC}(\textup{simple-}\Sigma^+_{t}[r])\equiv^{\fpt}p\textup{-MC}(\textup{simple-}\Sigma_{t}[r]) \equiv^{\fpt} p\textup{-MC}(\textup{simple-}\Sigma^-_{t}[r]),
\]
which yields part (3) by what we already have proven.

\qed

A reduction to the positive  (resp.\ negative) fragment is accomplished by:

\begin{lem}\label{lem:mcra}
\begin{enumerate}
\item
If $t\ge 1$ is odd, then $p\textup{-MC}(\Sigma_{t})\leq p\textup{-MC}(\Sigma^+_{t}).$
\item
If $t\ge 1$ is even, then $p\textup{-MC}(\Sigma_{t})\leq p\textup{-MC}(\Sigma^-_{t}).$
\end{enumerate}
\end{lem}

\proof Let $(\str{A},\phi)$ be an instance of $p\textup{-MC}(\Sigma_{t})$.
We may assume that all negation symbols in $\phi$ are in front of atomic subformulas. We
give an \fpt-reduction mapping $(\str{A},\phi)$ to a pair
$(\str{A}',\phi')$ with $(\str{A}\models \phi\iff \str{A}'\models
\phi') $, where $\phi'$ is a $\Sigma^+_{t}$-formula if $t$ is odd and a
$\Sigma^-_{t}$-formula if $t$ is even.

Let $\tau$ be the vocabulary of $\str{A}$.  The $\tau'$-structure $\str{A}'$
will be an expansion of $\str{A}$. It has an ordering $<^{\str{A}'}$ of its
universe $A'=A$. If $R\in\tau$ is $r$-ary, in $\tau'$ we have $r$-ary
relation  symbols $R_f$ and $R_l$, and a $2\cdot r$-ary relation symbol $R_s$.
$R_f^{\str{A}'}$ and $R_l^{\str{A}'}$ are singletons consisting of the first
and last tuple in $R^{\str{A}}$, respectively, in the lexicographic ordering
of $r$-tuples induced by $<^{\str{A}'}$ (and are empty in case
$R^{\str{A}}$ is empty). The relation $R_s^{\str{A}'}$ contains $(\bar a,\bar
b)$ iff ($R^{\str{A}}\bar a$, $R^{\str{A}}\bar b$, $\bar a$ is less than $\bar
b$, and no tuple in $R^{\str{A}}$ is between $\bar a$ and $\bar b$ in the
lexicographic ordering of $r$-tuples). Let $\bar y<_r\bar z$  denote a
quantifier-free formula of vocabulary $\tau'$ without the negation symbol
expressing that $\bar y$ is less than $ \bar z$ in the lexicographic ordering
of $r$-tuples.

Now assume that $t$ is odd. Then in
$\phi$ we replace every negative occurrence $\neg Rx_1\ldots x_r$ of $R$ by
\[
\exists y_1\ldots \exists y_r\exists z_1\ldots \exists z_r((R_f\bar y\land \mbox{$\bar x<_r\bar y$})\lor (R_s\bar y\bar z\land\mbox{$\bar y<_r\bar x$}\land\mbox{$\bar x<_r\bar z$})\lor (R_l\bar z\land\mbox{$\bar z<_r\bar x$})) 
\]
and every negative occurrence $\neg x=y$ by $(x<y\vee y<x)$.
The resulting formula is easily seen to be equivalent to a
$\Sigma^+_t$-formula $\phi'$.  If $t$ is even, we replace every positive
occurrence $Rx_1\ldots x_r$ of $R$ in $\phi$ by
\[
\neg\exists y_1\ldots \exists y_r\exists z_1\ldots \exists z_r((R_f\bar y\land \mbox{$\bar x<_r\bar y$})\lor (R_s\bar y\bar z\land\mbox{$\bar y<_r\bar x$}\land\mbox{$\bar x<_r\bar z$})\lor (R_l\bar z\land\mbox{$\bar z<_r\bar x$})) 
\]
and every positive occurrence $ x=y$ by $(\neg x<y\wedge \neg y<x)$.
We obtain a formula that is equivalent to a $\Sigma^-_t$-formula $\phi'$.
\qed

The last gap in a proof of Lemma~\ref{lem:mcr}(2), namely the transition to the simple fragments, will be closed by the following result:

 \begin{lem}\label{lem:sim}For   $t\ge 1$, 
$$
p\textup{-MC}(\Sigma_{t}[2])\leq p\textup{-MC}(\textup{simple-}\Sigma_{t}[3]).
$$
\end{lem}
\proof To simplify the notation we fix the parity of $t$, say, $t$ is even. Let  $(\str{A},\varphi)$ be an instance of $p\textup{-MC}(\Sigma_{t}[2])$. Thus, the
vocabulary $\tau$ of \str{A} has arity $\le 2$ and we can assume that the quantifier-free part of the sentence $\varphi$ is in conjunctive normal form,
\[
\varphi=\exists \bar y_1\forall \bar y_2\exists \bar y_3\ldots \forall \bar y_t\bigwedge_{i\in I}\bigvee_{j\in J} \lambda_{ij}
\]
with literals $\lambda_{ij}$. First we replace the conjunction $\bigwedge_{i\in
  I}$ in $\phi$ by a universal quantifier. For this purpose, we add to the
vocabulary $\tau$  unary relation symbols $R_i$ for $i\in I$ and
consider an expansion $\str{B}:=(\str{A},(R_i^{\str{B}})_{i\in I})$ of
\str{A}, where $(R_i^{\str{B}})_{i\in I}$ is a partition of $A$ into nonempty
disjoint sets. Then,
\[
\str{A}\models \varphi\iff \str{B}\models  \exists \bar y_1\forall \bar y_2\exists \bar y_3\ldots \forall \bar y_t\forall y\bigvee_{i\in I}\bigvee_{j\in J}(R_iy\wedge \lambda_{ij}).
\]
Since the arity of $\tau$ is $\le 2$, every $\lambda_{ij}$ contains at most two variables, say, $\lambda_{ij}=\lambda_{ij}(x_{ij},y_{ij})$. We expand \str{B} to a structure \str{C} by adding, for
all $i\in I$ and $j\in J$, a relation $T_{ij}^{\str{C}}$ of arity 3 
containing all triples $(a,b,c)$ such that $R_i^{\str{B}}a$ and $\str{B}\models \lambda_{ij}(b,c).$
Then,
\[
\str{A}\models \varphi\iff \str{C}\models \exists \bar y_1\forall \bar y_2\exists \bar y_3\ldots \forall \bar y_t\forall y \bigvee_{i\in I}\bigvee_{j\in J} T_{ij}yx_{ij}y_{ij}.
\]
The formula on the right hand side  is simple, so this equivalence yields the desired reduction.
\qed

\medskip\noindent \textit{Proof of Lemma~\ref{lem:mcr}(2):} By applying 
Lemma~\ref{lem:mcra}, Lemma~\ref{lem:bin}, Lemma~\ref{lem:sim}, and Lemma~\ref{lem:bin} one by one, we obtain the
following chain of reductions, say, for even $t$,
\[
p\textup{-MC}(\Sigma_{t})
\leq
p\textup{-MC}(\Sigma^-_{t})
\leq
p\textup{-MC}(\Sigma_{t}[2])
\le
p\textup{-MC}(\textup{simple-}\Sigma_{t}[3])
\le
p\textup{-MC}(\textup{simple-}\Sigma_{t}[2]).
\]
\qed 

When trying to prove Lemma~\ref{lem:mcr}(1) we are facing another difficulty:
For example, consider the case $t=3$. If we apply the reduction used
to prove Lemma~\ref{lem:mcr}(2)  to a formula in $\Sigma_{3,u}$,  the resulting
formula, even though equivalent to a formula in $\Sigma_3[2]$, is not
necessarily equivalent to a formula in $\Sigma_{3,u}[2]$.

The crucial property we exploit in our proof of Lemma~\ref{lem:mcr}(1) is that
in a $\Sigma_{t,u}$-formula the number of variables not occurring in the first,
existentially quantified, block of variables is bounded by $(t-1)\cdot u$. We
proceed in three steps: We start with $p\textup{-MC}(\Sigma_{t,u})$. In Lemma
\ref{lem:arb} we show how to pass from $\Sigma_{t,u}$ to $\Sigma_{t,u'}[r]$ for
some $u',r$; in Lemma \ref{lem:ari} we see that we can choose $r=2$. Finally, we
get $u'=1$ by Lemma \ref{lem:qua}.

 \begin{lem}\label{lem:arb} 
For  $t\geq 2$ and $u\ge 1$,
\[
p\textup{-MC}(\Sigma_{t,u})\leq p\textup{-MC}(\Sigma_{t,u+1}[t\cdot u]).
\]
\end{lem}

\proof Let $(\str{A},\phi)$ be an instance of $p\textup{-MC}(\Sigma_{t,u})$.
Say, $\phi=\exists x_1\ldots \exists x_k \psi$, where $\psi$ begins with a
universal quantifier. Set $q:=(t-1)\cdot u$ and let $\bar y=y_1,\ldots, y_q$
contain the variables in $\phi$ distinct from $x_1,\ldots, x_k$. We shall define a
structure $\str{A}'$ and a $\Sigma_{t,u+1}[t\cdot u]$-sentence $\phi'$ with
$(\str{A}\models \phi\iff \str{A'}\models \phi')$. 

Let $\Lambda$ be the set of all atomic subformulas of $\phi$. Here  the
notation $\lambda(x_{i_1},\ldots,x_{i_\ell},\bar y)$ indicates that $x_{i_1},\ldots,x_{i_\ell}$ are the variables from $x_1,\ldots, x_k$ in $\lambda$. The vocabulary $\tau'$ of $\str{A}'$ contains a unary relation symbol $O$ (the ``old element relation''),
binary relation symbols $E_1,\ldots, E_k$ (the ``component relations'') and
for every $\lambda(x_{i_1},\ldots, x_{i_{\ell}},\bar y)\in\Lambda$ a unary
relation symbol $W_{\lambda}$ and a $(1+q)$-ary relation symbol $R_{\lambda}$.
Thus the arity of $\tau'$ is at most $1+q\le t\cdot u$. For every
$\lambda\in\Lambda$ and
$a_1,\ldots, a_{\ell}\in A$ with
\begin{equation}\label{equ:wit}\str{A}\models \exists \bar y\lambda(a_1,\ldots, a_{\ell},\bar y)
\end{equation}
we have in $A'$ a new element $w(\lambda,a_1,\ldots, a_{\ell})$, 
a ``witness'' for (\ref{equ:wit}). We let
\begin{align*} 
A'&:=A\cup\big\{w(\lambda,a_1,\ldots, a_{\ell}) 
\bigmid
  \lambda(x_{i_1},\ldots, x_{i_{\ell}},\bar y)\in\Lambda,\,
\bar   a=(a_1,\ldots,a_\ell)\in A^{\ell},\, 
\str{A}\models \exists \bar y\lambda(\bar a,\bar y)\big\},\\
O^{\str{A}'}&:=A\\
E_i^{\str{A}'}&:=\big\{(a_i,w(\lambda,a_1,\ldots, a_{\ell})) \bigmid 
w(\lambda,a_1,\ldots, a_{\ell})\in A' \big\}\quad(\text{for }1\le i\le k).
\end{align*}
For every $\lambda\in\Lambda$ we let:
\begin{align*}
W_{\lambda}^{\str{A}'}&:=
\big\{w(\lambda,a_1,\ldots, a_{\ell}) \bigmid \bar a\in A^{\ell}\text{ and }\str{A}\models \exists \bar y\lambda(\bar a,\bar y)\big\},\\
R_{\lambda}^{\str{A}'}&:=
\big\{\big(w(\lambda,a_1,\ldots, a_{\ell}), b_1,\ldots, b_q\big) \bigmid  \bar a\in A^{\ell},\, \bar b\in A^{q}, \text{ and }\str{A}\models \lambda(\bar a,\bar b)\big\}.
\end{align*}
This completes the definition of $\str{A}'$. Note that $\|\str{A}'\|\leq O(\|\str{A}\|^q\cdot |\phi|)$. 

For every $\lambda(x_{i_1},\ldots, x_{i_{\ell}},\bar y)\in\Lambda$ let
$\chi_{\lambda}(x_{i_1},\ldots, x_{i_{\ell}},z_{\lambda})$ be a formula
expressing:
\begin{quote}
``Either $z_{\lambda}\in W_{\lambda}$ is the witness  for $x_{i_1},\ldots,
x_{i_{\ell}}$  or $z_{\lambda}\notin W_{\lambda}$ and there is no witness in
$W_{\lambda}$ for $x_{i_1},\ldots, x_{i_{\ell}}$.''
\end{quote}
That is, we let
\[\chi_{\lambda}(x_{i_1},\ldots, x_{i_{\ell}},z_{\lambda})\, :=\, 
(W_{\lambda} z_{\lambda}\land\bigwedge_{j=1}^{\ell} E_j x_{i_j}z_{\lambda})\lor (\neg W_{\lambda} z_{\lambda}\wedge\forall y ¬(W_{\lambda} y\land\bigwedge_{j=1}^{\ell} E_j x_{i_j}y)).
\]
Then, for $\bar a\in A^{\ell},\bar b\in A^q$, and $c\in A'$, we have:
\begin{center}
  If $\str{A}'\models \chi_{\lambda}(\bar a, c)$ then $(\str{A}\models
  \lambda(\bar a,\bar b)\iff\str{A}'\models R_{\lambda}z_\lambda\bar y(c\bar
  b)) $.
\end{center}
Let $\chi:=\bigwedge_{\lambda\in\Lambda}\chi_{\lambda}$.
Let $\psi'$ be the formula obtained from $\psi$ by replacing every atomic subformula $\lambda(x_{i_1},\ldots, x_{i_{\ell}},\bar y)$ by $R_{\lambda}z_{\lambda}\bar y$ and relativizing all quantifiers to $O$. Finally, we let
\[
\phi':=\exists x_1\ldots \exists x_k\exists (z_{\lambda})_{\lambda\in\Lambda}(Ox_1\wedge\ldots \wedge Ox_k\wedge\psi'\land \chi).
\]
Then
\[
\str{A}\models \phi\iff \str{A'}\models \phi'.
\] 
Since $\chi$ is equivalent to a formula of the form $\forall z \chi'$ with
quantifier-free $\chi'$, the quantifier $\forall z$ can be added to the first
block of $\psi'$ (recall that $t\ge 2$). Thus,  the formula $\varphi'$ is equivalent to a formula in 
$\Sigma_{t,u+1}[t\cdot u]$.
\qed

\begin{lem}\label{lem:ari}For  $t\ge 2$ and $u,r\geq 1$,
\[
p\textup{-MC}(\Sigma_{t,u}[r])\leq p\textup{-MC}(\textup{strict-}\Sigma_{t,u+1}[2]).
\]
\end{lem}

\proof
Let $(\str{A},\phi)$ be an instance of
$p\textup{-MC}(\Sigma_{t,u}[r])$. 
We shall define a structure $\str{A}'$ of vocabulary $\tau'$ of arity 2 and a
strict-$\Sigma_{t,u+1}$-sentence $\phi'$ such that ($\str{A}\models \phi\iff
\str{A}'\models \phi'$). 

For notational simplicity, let us assume that $t\ge 2$ is even. Suppose that 
\[
\phi=\exists x_1\ldots\exists x_k\forall\bar y_1\exists\bar y_2\ldots \forall\bar
y_{t-1}\psi,
\]
where $\psi$ is quantifier-free and $\bar y_i=(y_{(i-1)u+1},\ldots,y_{iu})$. Let
$\bar y=(y_1,\ldots, y_{(t-1)u})$. Let $\Lambda$ be the
set of all atomic subformulas of $\phi$, $\tau$ the
vocabulary of $\mathcal A$, and $r_0:=\max\{r,(t-1)\cdot u \}$.

The vocabulary $\tau'$ contains the unary relations symbols $T_1,\ldots, T_{r_0}$,
the binary relation symbols $E_1,\ldots,E_{r_0}$, and a binary relation symbol
$S_\lambda$ for every $\lambda\in\Lambda$.

 The universe of the structure $\mathcal A'$ is  $A':=A\cup
 A^2\cup\ldots \cup A^{r_0}$. The relation symbols are interpreted as follows:
\begin{itemize}
\item
For $1\le i\le r_0$, \ \  $T_i^{\str{A}'}=A^i$.
\item
For $1\le i\le r_0$, 
\[
E_i^{\str{A}'}:=\{(a_i,(a_1,\ldots, a_s)) \mid   i\leq s\le r_0,\, (a_1,\ldots, a_s)\in A^s \}.
\]
\item
For every $\lambda(x_{i_1},\ldots,x_{i_s},\bar
y)\in\Lambda$
we let
\[
S_\lambda^{\mathcal A'}:=\big\{(\bar a,\bar b)\bigmid \bar
a\in A^s,\bar b\in A^{(t-1)u},\text{ and }\mathcal A\models\lambda(\bar
a,\bar b)\big\}.
\]
\end{itemize}
Note that the size of $\mathcal A'$ is $O(|A|^{r_0(t-1)u})$ and thus
polynomial in the size of $\mathcal A$.

To define the formula $\phi'$, for every $\lambda(x_{i_1},\ldots,x_{i_s},\bar
y)\in\Lambda$  we introduce a new variable $x_\lambda$ and let
\[
\chi_\lambda(\bar x,x_\lambda):=T_sx_\lambda\wedge E_1x_{i_1}x_\lambda \wedge\ldots\wedge
E_sx_{i_s}x_\lambda.
\]
Furthermore, we let $\chi=\bigwedge_{\lambda\in\Lambda}\chi_\lambda$.
We introduce another new variable $y$ representing the whole tuple $\bar y$
and let
\[
\xi(\bar y,y):=T_{(t-1)u}y\wedge\bigwedge_{i=1}^{(t-1)u} E_iy_iy.
\]
Finally, we let 
\[
\eta(v_1,\ldots, v_u):= T_1v_1\wedge\ldots\wedge T_1 v_u
\]
and let $\phi''$ be the formula
\begin{align*}
&\exists x_1\ldots\exists x_k\exists (x_\lambda)_{\lambda\in\Lambda}
\Big(\chi
\wedge\\
&\hspace{1cm}\forall\bar y_1\big(\eta(\bar y_1)\to\exists\bar y_2(\eta(\bar y_2)\wedge \ldots\wedge\forall \bar y_{t-1}(\eta(\bar y_{t-1})\to\forall y
(\xi(\bar y,y)\to\psi'))\ldots )\big)\Big),
\end{align*}
where $\psi'$ is the formula obtained from $\psi$ by replacing each
$\lambda\in\Lambda$ by the atom $S_\lambda x_\lambda y$.
It is easy to see that ($\str{A}\models \phi\iff
\str{A}'\models \phi''$) and that $\phi''$ is equivalent to a formula in
$\Sigma_{t,u+1}[2]$. 

However, it is not obvious how to translate $\phi''$ to a formula in 
$\text{strict-}\Sigma_{t,u+1}[2]$. The problematic atoms are those of the form
$E_ix_j x_\lambda$ in the formula $\chi$. 
To resolve the problem, we introduce a new variable $z$ and let, for $\lambda(x_{i_1},\ldots,x_{i_s},\bar y)\in\Lambda$, 
\[
\chi_\lambda'(\bar x,x_\lambda)=T_sx_\lambda\wedge \forall z(z=x_\lambda\to
E_1x_{i_1}z\wedge\ldots\wedge
E_sx_{i_s}z).
\]
We let $\chi'=\bigwedge_{\lambda\in\Lambda}\chi_\lambda'$ and $\phi'''$ the
formula obtained from $\phi''$ by replacing the subformula $\chi$ by
$\chi'$. It is easy to transform $\phi'''$ into a formula in
$\textup{strict-}\Sigma_{t,u+1}[2]$.
\qed

The following lemma, the last step of our proof, is a the ``strict version''
of a result of \cite{flugro01}. For the reader's convenience, we sketch the
simple proof:

\begin{lem}\label{lem:qua}For   $t\ge 2$ and  $u,r\geq 1$,
\[
p\textup{-MC}(\textup{strict-}\Sigma_{t,u}[r])
\leq 
p\textup{-MC}(\textup{strict-}\Sigma_{t,1}[r]).
\]
\end{lem}

\noindent 
\textit{Proof (sketch):}
For simplicity we let $t=3$. Let $(\str{A},\phi)$ be an instance of
$p\textup{-MC}(\textup{strict-}\Sigma_{3,u}[r])$. Let $\tau $ be the
vocabulary of $\str{A}.$ The sentence $\phi$ has the form
\[
\exists x_1\ldots \exists x_k\forall \bar y\exists \bar z \psi
\]
with quantifier-free $\psi$ and with $|\bar y|=|\bar z|=u$. We shall construct an
equivalent instance $(\str{A}',\phi')$ of
$p\textup{-MC}(\textup{strict-}\Sigma_{t,1}[r])$.

We set $A':=A\cup A^u$. The new unary relation symbol $T$ is interpreted in
$\str{A}'$ by $T^{\str{A}'}:=A^u$. For every atomic subformula $\lambda$ of
$\phi$, say $\lambda=Rx_2y_3z_6y_4x_2$, we introduce a new relation symbol
$R_{\lambda}$ and set
\begin{eqnarray*}R_{\lambda}^{\str{A}'}abc&\mbox{iff}& a\in A, \, b,c\in A^u\, \mbox{\ and\ } R^{\str{A}}ab_3c_6b_4a, \mbox{\ where $b_3,b_4$, and $c_6$ are} \\
&&\mbox{the third and fourth component of $b$ and the sixth component of $c$, respectively.}
\end{eqnarray*} 
Finally, we set $\phi':=\exists x_1\ldots \exists x_k\forall  y\exists  z (\bigwedge_{1\le i \le k}\neg Tx_i\wedge( Ty\to( Tz\wedge \psi')))$, where $\psi'$ is obtained from $\psi$ by replacing any atomic subformula $\lambda=Rx_2y_3z_6y_4x_2$ by $R_{\lambda}x_2yz$. \qed

\medskip\noindent \textit{Proof of Lemma~\ref{lem:mcr}(1): }Combining
Lemmas~\ref{lem:arb}, \ref{lem:ari}, and \ref{lem:qua}, we obtain the
following chain of reductions:
\[
p\textup{-MC}(\Sigma_{t,u})
\leq 
p\textup{-MC}(\Sigma_{t,u+1}[t\cdot u])
\le
p\textup{-MC}(\textup{strict-}\Sigma_{t,u+2}[2])
\le 
p\textup{-MC}(\textup{strict-}\Sigma_{t,1}[2]).
\]
\qed

\begin{rem}\label{rem:graph}
The First-Order Normalisation Lemma shows that for the model-checking problems
for the various classes of first-order formulas we are interested in, it
suffices to consider binary vocabularies. However, these vocabularies may
still contain arbitrarily many unary and binary relation symbols. We can
further strengthen the results to vocabularies with just one
binary relation symbol and also restrict the input structures in the
model-checking problems to be (simple undirected) graphs. 

For every class $\Phi$ of formulas we consider the following restriction of
 $p\textup{-MC($\Phi$[2])}$:

\npprob{$p\textup{-MC($\Phi$[GRAPH])}$ }{A graph \str{G}}{A sentence $\phi$
  in $\Phi$}{Decide if \str{G} satisfies $\phi$}

The following strengthening of Lemma~\ref{lem:mcr}(3) is already proved in
\cite{flugro01}:
\begin{enumerate}
\item[($3'$)] 
$p\textup{-MC}(\FO)\le p\textup{-MC}(\FO[\text{GRAPH}]).$
\end{enumerate}
Furthermore, it is proved in \cite{flugro01} that for every $t\ge 1$,
\[
p\textup{-MC}(\Sigma_{t}[2])\le p\textup{-MC}(\Sigma_{t}[\text{GRAPH}]).
\]
Together with  Lemma~\ref{lem:mcr}(2) this yields
\begin{enumerate}
\item[($2'$)] 
For all $t\ge 1$, $p\textup{-MC}(\Sigma_t)\le p\textup{-MC}(\Sigma_t[\text{GRAPH}]).$
\end{enumerate}
The corresponding strengthening of (1) is not so obvious, and we still do not know a direct
proof. Surprisingly, the result can be shown by taking a detour via
propositional logic, as we will see in the next section
(cf.~Corollary~\ref{cor:gra}).
\end{rem}

\section{Back and forth between propositional and first-order logic: 
  the basic machinery}\label{sec:bac}

In their most basic form, the results of this section go back to Downey,
Fellows, and Regan \cite{dowfelreg98}. We have (slightly) improved these
results in an earlier paper \cite{flugro01}, and here we give another
improvement. Moreover, we present a new proof, which we believe is significantly simpler
than those known for the weaker versions of the theorems. The proofs of all
results presented in the later sections of this paper are based on the ideas
developed here.

As an application, we show some of the core results of Downey and Fellows
structure theory for the W-hierarchy; in particular, the ``Normalisation
Theorem'' and its sharpened version for monotone/anti-monotone formulas
(cf.~Chapter~12 of \cite{dowfel99}) are  easy corollaries of
Theorem~\ref{the:afo}.

\subsection{From propositional to first-order logic}\label{sub:fro}
In this subsection we show how to reduce weighted satisfiability problems for
propositional formulas to model-checking problems for fragments of first-order
logic. For this purpose we need a known algorithm computing minimal covers in
hypergraphs. We recall the fact.

Let $\str{H}=(H,E)$ be a {\em hypergraph}, i.e., $H$ is a set, the set of {\em
  points} of $\str{H}$, and $E$ is a set of non-empty subsets of $H$, the set
of {\em edges} of $\str{H}$. A subset $X\se H$ {\em covers} an edge $e\in E$,
if $X\cap e\not=\emptyset$; $X$ covers \str{H}, if $X$ covers all edges of
\str{H}. If $X$, but no proper subset of $X$, covers \str{H}, then $X$ is a
{\em minimal cover} of \str{H}. The \emph{arity} of a hypergraph
 is the maximum cardinality of its edges. 

\begin{lem}\label{cover}There is an algorithm that, given a hypergraph $\mathcal H$ of
  arity at most $d$,
  computes in time $O(k\cdot d^k\cdot \|\str{H}\|)$ a list of all minimal covers of\/
  $\mathcal H$ of size at most $k$.
\end{lem}

\proof The algorithm is a straightforward generalisation of a standard
algorithm (using the bounded search tree technique, cf.~\cite{dowfel99}) showing that the parameterized vertex cover problem is in \FPT.

Let $\str{H}=(H,E)$ be as in the statement of the lemma. Let $e_1,\ldots, e_m$
be an enumeration of $E$. The algorithm builds a labelled $d$-ary tree of
depth $\le k$. The labels of the nodes are pairs $(X,i)$, where $X\subseteq H$
with $|X|\le k$ and $0\le i\le m$. Label $(X,i)$ gives
the information that $X$ covers the edges $e_1,\ldots,e_i$, but not $e_{i+1}$
(if $i+1\le m$).

The construction of the tree is by induction: The label of the root is
$(\emptyset,0)$. Suppose that a node $t$ is labelled by $(X,i)$. If $i=m$ or if
the depth of $t$ is $k$, then $t$ has no child.  Otherwise, let
$e_{i+1}=\{h_1,\ldots,h_s\}$. Node $t$ has children $t_1,\ldots,t_s$. For
$1\le j\le s$, the label of $t_j$ is $(X_j,i_j)$, where $X_j=X\cup\{h_j\}$ and
 $i_j$ is the maximum index such that $X_j$ covers $e_1,\ldots,e_{i_j}$.

One easily verifies that any set $Y$ of at most $k$ points is a cover of \str{H} if and only if there is a leaf of the tree labelled by $(X,m)$ with $X\se Y$. Thus, to obtain the list of all minimal covers of size $\le k$, the algorithm checks for every leaf labelled by $(X,m)$, whether $X$ is a minimal cover. For this purpose, it simply tests for each of the at most $k$ subsets obtained by
removing a single element from $X$ if it is a cover.

\qed 

In a first step (Lemma \ref{lem:g1d}) we give the translation of formulas in
$\Gamma_{1,d}$ to first-order logic. Recall that a set $\{X_1,\ldots,X_k\}$ of
propositional variables represents the assignment that sets $X_1,\ldots,X_k$ to
\textsc{true} and all other variables to \textsc{false}.

\begin{lem}\label{lem:g1d}
   For all $d,k\ge 1$ and for all formulas
  $\alpha(X_1,\ldots,X_m)\in\Gamma_{1,d}$ there are
  \begin{itemize}
  \item a structure  $\str A=\str A_{\bigwedge,\alpha(X_1,\ldots,X_m),d,k}$ with universe $A:=\{1,\ldots,
    m \}$,
\item a quantifier-free formula $\psi=\psi_{\bigwedge,d,k}(x_1,\ldots,x_k)$ that only depends
  on $d$ and $k$, but not on $\alpha$
\end{itemize}
such that the mapping $(\alpha,d,k)\mapsto(\mathcal A,\psi)$ is computable in time $k^{d+2}\cdot d^k\cdot p(|\alpha|)$ for some polynomial
  $p$  and such that
for  pairwise distinct $m_1,\ldots,m_k\in A$
\[
\{X_{m_1},\ldots, X_{m_k}\}\text{ satisfies }\alpha\iff \str{A}\models \psi(m_1,\ldots, m_k).
\]
\end{lem}

\proof Let $d\ge 1$, $\alpha(X_1,\ldots, X_m)\in \Gamma_{1,d}$, and $k\in
\mathbb N$ be given, say,
\[
\alpha=\bigwedge_{i\in I}\delta_i, \mbox{\ \quad where each $\delta_i$ is the disjunction of $\leq d$ literals.}
\]
We may assume that every $\delta_i$ has the form
\begin{equation}\label{equ:dis}
¬ X_{i_1}\lor\ldots \lor¬ X_{i_r}\lor X_{j_1}\lor\ldots \lor X_{j_s} 
\end{equation}
with $0\le r,s$ and $1\le r+s\leq d$ and with pairwise distinct $X_{i_1},\ldots, X_{j_s}$.

We call $t:=(r,s)$ the {\em type} of  $\delta_i$. The structure \str{A}  has universe $A:=\{1,\ldots, m \}$; for every type  $t=(r,s)$, the structure \str{A} contains the $r$-ary relation 
\[
V_t^{\str{A}}:=\big\{(i_1,\ldots, i_r) \bigmid\text{there are
  $j_1,\ldots,j_s$ such that clause (\ref{equ:dis}) occurs in
  $\alpha\big\}$.}
\] 
The structure $\str{A}$ contains further relations that will be defined later.

The formula $\psi$ will have the form $\bigwedge_{t\textup{\ 
    type}}\psi_t$, where $\psi_t=\psi_t(x_1,\ldots, x_k)$ will express in
\str{A} that $X_{x_1},\ldots, X_{x_k}$ satisfies every clause of type $t$
of $\alpha$.  If $t=(r,0)$ set
\[
\psi_t:=\bigwedge_{1\leq i_1,\ldots,i_r\leq k}¬ V_t x_{i_1}\ldots  x_{i_r}.
\]
Let $t=(r,s)$ with $s\not= 0$. Fix $(i_1,\ldots, i_r)\in V_t^{\str{A}}$. Then, for $1\le m_1,\ldots, m_k\le m$,
\begin{quote}\sloppy the assignment $X_{m_1},\ldots,X_{m_k} $ satisfies all clauses $¬ X_{i_1}\lor\ldots \lor¬ X_{i_r}\lor X_{j_1}\lor\ldots \lor X_{j_s}$ in $\alpha$\\[2mm] 
\centerline{if and only if}\\[2mm] $X_{m_1},\ldots,X_{m_k} $ satisfies
$¬ X_{i_1}\lor\ldots \lor¬ X_{i_r}$ or $\{m_1,\ldots, m_k \}$ is a
cover of the hypergraph $\str{H}\, (=\str{H}_t(i_1,\ldots, i_r))\,
:=(H,E)$ with
\begin{align*}
H&:=\{1,\ldots, m  \}\mbox{ and }\\
E&:=\{\{j_1,\ldots, j_s  \} \mid ¬ X_{i_1}\lor\ldots \lor¬ X_{i_r}\lor X_{j_1}\lor\ldots \lor X_{j_s} \mbox{\ occurs in $\alpha$}\}.
\end{align*}
\end{quote}
Let $C_1,\ldots, C_{d^k}$ be an enumeration (with repetitions if necessary) of the minimal covers of \str{H} of size $\leq k$. View every $C_i$ as a sequence of length $k$ (with repetitions if necessary). For $u=1,\ldots, d^k$ and $\ell=1,\ldots, k$ add to $\str{A}$ the $(r+1)$-ary relations $L_{t,u,\ell}^{\str{A}}$, where
\[
L_{t,u,\ell}^{\str{A}}:=\{(i_1,\ldots, i_r,v) \mid \mbox{$v$ is the $\ell$th element of the $u$th cover $C_u$ of $\str{H}_t(i_1,\ldots, i_r)$}\}
\]
(if $\str{H}_t(i_1,\ldots, i_r)$ has no cover of size $\le k$, then $L_{r,u,\ell}^{\str{A}}$ contains no tuple of the form $(i_1,\ldots, i_r,v)$). Now the preceding equivalence shows that we  can set
\[
\psi_t:=\bigwedge_{1\leq i_1,\ldots , i_r\leq k}(V_t x_{i_1}\ldots x_{i_r}\to\bigvee_{1\leq u\leq d^k}\bigwedge_{1\leq \ell\leq k}\bigvee_{1\leq j\leq k}L_{t,u,\ell} x_{i_1}\ldots  x_{i_r}x_j).
\]
It is easy to see that $\mathcal A$ and $\psi$ can be computed from $\alpha$, $d$,
and $k$ in time $k^{d+2}\cdot d^k\cdot p(\alpha)$ for some polynomial $p$; the only nontrivial part is the
computation of the list of minimal covers, which is taken care of by  Lemma \ref{cover}.
\qed

\begin{cor}\label{cor:dis}
  For all $d,k\ge 1$ and for all formulas
  $\alpha(X_1,\ldots,X_m)\in\Delta_{1,d}$ there are
  \begin{itemize}
  \item a structure  $\str A=\str A_{\bigvee,\alpha(X_1,\ldots,X_m),d,k}$ with universe $A:=\{1,\ldots,
    m \}$,
\item a quantifier-free formula $\psi=\psi_{\bigvee,d,k}(x_1,\ldots,x_k)$ that only depends
  on $d$ and $k$, but not on $\alpha$
\end{itemize}
such that the mapping $(\alpha,d,k)\mapsto(\mathcal A,\psi)$ is computable in time $k^{d+2}\cdot d^k\cdot p(|\alpha|)$ for some polynomial
  $p$ and 
for  pairwise distinct $m_1,\ldots,m_k\in A$
\[
\{X_{m_1},\ldots, X_{m_k}\}\text{ satisfies }\alpha\iff \str{A}\models \psi(m_1,\ldots, m_k).
\]
\end{cor}

\proof Exploiting the fact that $\neg\alpha$ is equivalent to a formula
$\alpha'$ in $\Gamma_{1,d}$, we let $\str A$ be the structure constructed in
Lemma~\ref{lem:g1d} for the formula $\alpha'$ and
$\psi_{\bigvee,d,k}:=\neg\psi_{\bigwedge,d,k}$.  \qed

\begin{cor}$\textsc{WSat}(\Gamma_{1,d}\cup \Delta_{1,d})\leq p\textup{-MC}(\Sigma_1)$.
\end{cor}

\proof Given an instance $(\alpha,k)$ of $\textsc{WSat}(\Gamma_{1,d}\cup
\Delta_{1,d})$, compute $(\mathcal A,\psi)$ as in Lemma~\ref{lem:g1d} or
Corollary~\ref{cor:dis}. Let 
\[
\phi=\exists x_1\ldots\exists x_k\big(\bigwedge_{1\le i<j\le k}x_i\neq
x_j\wedge\psi\big).
\]
Then 
\[
\alpha \mbox{\ is $k$-satisfiable}\iff \str{A}\models\phi,
\]
which gives the desired reduction.
\qed

Lemma \ref{lem:g1d} and Corollary \ref{cor:dis} show how to translate formulas
in $\Gamma_{1,d}\cup\Delta_{1,d}$ to quantifier-free formulas. When
translating propositional formulas of a weighted satisfiability problem into
first-order formulas of a model-checking problem, every additional big
conjunction and big disjunction leads to a universal and  an existential
quantifier, respectively. The following proposition is based on this
observation.

\begin{prop}\label{pro:afo} 
For  all $d,t\geq 1$ 
\[
\textsc{WSat}(\Delta_{t+1,d})\leq p\textup{-MC}(\Sigma_{t,1}).
\] 
\end{prop} 
\proof Fix $d,t\geq 1$. 
Let $(\alpha,k)$ be an instance of
$\textsc{WSat}(\Delta_{t+1,d})$.
We shall construct a structure $\mathcal
A$ and a $\Sigma_{t,1}$-sentence $\phi$ such that
\begin{equation}\label{equa:w}
\alpha\text{ is $k$-satisfiable}\iff \mathcal A\models \phi. 
\end{equation}

Let the variables of $\alpha$ be $X_1,\ldots,X_m$. We assume that $t$ is even, the case ``$t$ is odd'' is handled analogously. Thus $\alpha$ is of the form
\[\bigvee_{i_1\in I_1}\bigwedge_{i_2\in I_{i_1}}\ldots\bigwedge_{i_t\in I_{i_1\ldots i_{t-1}}}\delta_{(i_1,\ldots,i_t)},
\]
where the $\delta_{\bar{i}}\in\Delta_{1,d}$.
A simple argument shows that we can pass to an equivalent formula $\alpha'$ with $|\alpha'|\le |\alpha|^t$ of the form
\[
\bigvee_{i_1\in I_1}\bigwedge_{i_2\in I_2}\ldots\bigwedge_{i_t\in I_t}\delta_{(i_1,\ldots, i_t)},
\]
so we assume that $\alpha$ itself already has this form.
  Let $\bar I:=I_1\times\ldots\times I_t$. 

The structure $\mathcal A$ consists of two parts: The first part is the tree $\mathcal T$ of height $t$ obtained from the ``parse tree'' of $\alpha$ by removing all nodes that correspond to small subformulas of $\alpha$.  The edge relation of this tree, directed from the root to the leaves (which by definition have height 0), is represented by the binary relation $E^{\str A}$. Moreover, we add a unary relation symbol $\textup{Root}$ and let $\textup{Root}^{\str A}$ be the singleton containing the root of $\mathcal T$.  Note that each leaf of $\str T$ corresponds to a subformula $\delta_{\bar{i}}$, for some $\bar{i}\in\bar I$, of $\alpha$. We denote the leaf corresponding to $\delta_{\bar{i}}$ by $\ell_{\bar{i}}$. Each node of $\str T$ of height $s$ corresponds to a subformula contained in $\Gamma_{s+1,d}$ if $s$ is odd or $\Delta_{s+1,d}$ if $s$ is even.

The universe of the second part of $\mathcal A$ is $\{1,\ldots,m\}$ (the set of indices of the variables of $\alpha$). For every $\bar{i}\in\bar I$, let $\mathcal A_{\bar {i}}=\mathcal A_{\bigvee,\delta_{\bar {i}}(X_1,\ldots, X_m),d,k}$ be the structure defined in Corollary~\ref{cor:dis}. Essentially, the second part of $\str A$ simply consists of all $\mathcal A_{\bar {i}}$s. However, all $\mathcal A_{\bar {i}}$s have the same universe $\{1,\ldots,m\}$. To keep them apart, we ``tag'' the tuples belonging to a relation in $\mathcal A_{\bar {i}}$ with the leaf $\ell_{\bar {i}}$ of $\str T$ that corresponds to $\delta_{\bar{i}}$.  More precisely, for each $r$-ary relation symbol $R$ in the vocabulary of the $\mathcal A_{\bar {i}}$s, the vocabulary of $\mathcal A$ contains an $(r+1)$-ary relation symbol $R'$.  We let
\[
(R')^{\str A}:=\big\{(\ell_{\bar {i}},a_1,\ldots,a_r)\bigmid
\bar{i}\in\bar I,\,\,(a_1,\ldots,a_r)\in R^{\str A_{\bar{i}}}\big\}.
\]
Finally, to be able to tell the two parts of $\mathcal A$ apart, we add one
unary relation symbol $V$ and let $V^{\str A}:=\{1,\ldots,m\}$. This completes
the definition of $\mathcal A$.

We now define, by induction on $s\ge 0$, formulas $\psi_s(y,x_1,\ldots,x_k)$
such that for every node $b$ of $\str T$ of height $s$, corresponding to a
subformula $\beta$ of $\alpha$, and all pairwise distinct $a_1,\ldots,a_k\in\{1,\ldots,m\}$ we
have
\begin{equation}\label{eq:afo1}
\{X_{a_1},\ldots,X_{a_k}\}\text{ satisfies }\beta\iff\str
A\models\psi_s(b,a_1,\ldots,a_k).
\end{equation}
$\psi_0(y,x_1,\ldots,x_k)$ is the formula obtained from the formula $\psi_{\bigvee,d,k}$ of Corollary~\ref{cor:dis} by replacing each atomic
subformula $Rx_1\ldots x_r$ by $R'yx_1\ldots x_r$. Then for $s=0$, 
\eqref{eq:afo1} follows from our construction of $\str A$ and
Corollary~\ref{cor:dis}.

For even  $s\ge 0$, we  let 
\[
\psi_{s+1}(y,x_1,\ldots,x_k):=\forall
z\big(Eyz\to\psi_{s}(z,x_1,\ldots,x_k)\big),
\]
and \eqref{eq:afo1} follows from the fact that all nodes of height $s+1$
correspond to conjunction of formulas corresponding to nodes of height
$s$. Similarly, for odd $s\ge 0$ we let 
\[
\psi_{s+1}(y,x_1,\ldots,x_k):=\exists
z\big(Eyz\wedge\psi_{s}(z,x_1,\ldots,x_k)\big).
\]
Finally, we let
\[
\phi:=\exists x_1\ldots\exists x_k\exists y\Big(\bigwedge_{i=1}^k
Vx_i\wedge\bigwedge_{\substack{i,j=1\\i\neq j}}^kx_i\neq x_j\wedge\textup{Root}\, y\wedge\psi_{t}(y,x_1,\ldots,x_k)\Big).
\]
It is easy to see that $\phi$ is equivalent to a formula in $\Sigma_{t,1}$.
\qed

We consider a more general weighted satisfiability problem, in which the depth of the formula is not fixed but treated as a parameter. The preceding proof yields:

\begin{cor}\label{cor:gdk}
  $P\leq p\textup{-MC}(\Sigma_{t,1})$, where $P$ is the parameterized
  problem \npprob{ }{$k\in\mathbb N$ and
    $\alpha\in\Delta_{t+1,k}$}{$k$.}{Decide if $\alpha$ is
    $k$-satisfiable}
\end{cor}

For later reference, let us state the following lemma, which is an immediate
consequence of the preceding proof:

\begin{lem}\label{lem:afo}
  Let $t,d\ge 1$. Then for all $k\ge 1$ and for all formulas
  $\alpha(X_1,\ldots, X_m)\in\Delta_{t+1,d}$ there are
  \begin{itemize}
  \item a structure $\str A$ with a unary relation $V^{\mathcal A}=\{1,\ldots,
    m \}$,
  \item a formula $\psi(x_1,\ldots,x_k)$ of the form $\exists y\exists
    y_1\forall y_2\exists y_3\ldots Q_{t}y_{t}\psi'$, where $Q_{t}=\exists$ if
    $t$ is odd and $Q_{t}=\forall$ if $t$ is even and $\psi'$ is quantifier
    free, and the formula $\psi$ only depends on $t,d,k$, but not on $\alpha$,
  \end{itemize}
  such that the mapping $(\alpha,k)\mapsto(\str A,\psi)$ is fixed-parameter
  tractable and for pairwise distinct ${m_1},\ldots,m_k\in V^{\str A}$, 
  \[
  \{X_{m_1},\ldots, X_{m_k}\}\text{ satisfies }\alpha\iff\str{A}\models \psi(m_1,\ldots, m_k).
  \]
\end{lem}
Together with Lemma~\ref{lem:pn1}, Proposition~\ref{pro:afo}  immediately
yields:

\begin{cor}\label{cor:del}For all $d\ge t \ge 1$   we have
\[
\textsc{WSat}(\Omega_{t,d})\leq p\textup{-MC}(\Sigma_{t,1}).
\] 
\end{cor}
%

\subsection{From first-order to propositional logic}\label{sub:fr2}
We turn to a reduction from model-checking problems for the fragments
$\Sigma_{t,u}$ to weighted satisfiability problems for propositional formulas.
We shall see that single quantifiers (or blocks of quantifiers of bounded
length) translate into big disjunctions and conjunctions; the leading
unbounded block yields the propositional variables, and its length yields the
parameter.

We start by collecting some simple facts.

Let $A$ be a set and $k\ge 1$. For all $a\in A$ and $1\le i\le k$, let
$X_{i,a}$ be a propositional variable with $X_{i,a}\not=X_{j,b}$ for
$(i,a)\not=(j,b)$.  Let $V$ be the set of all these propositional variables.
Let us call an assignment $S\in 2^V$ \emph{functional} if for each $i$ there
is exactly one $a$ such that $X_{i,a}$ is \textsc{true}. The proof of the
following lemma is straightforward.
\begin{lem}\label{lem:mod}Let $V=\{X_{i,a} \mid 1\le i\le k,\ a\in A\}$.
\begin{enumerate}
\item For
\begin{eqnarray*}
\chi^-:=\bigwedge_{\substack{1\le i\le k\\a,b\in A, a\neq b}}
(\neg X_{i,a}\vee\neg X_{i,b})&\mbox{and}&
\chi^+:=\bigwedge_{i=1}^k\bigvee_{b\in A}X_{i,b}
\end{eqnarray*}
and for every assignment $S\subseteq V$ of
weight $|S|=k$ we have
\[
S\text{ satisfies }\chi^-
\iff S\text{ is functional}\iff S\text{ satisfies }\chi^+.
\]
Observe that  $\chi^-\in\Gamma_{1,2}^-$  and  $\chi^+\in\Gamma_{2,1}^+$. In addition, we may as well consider $\chi^-$ as a formula in 
$\Gamma_{2,1}^-$.
\item Let \str{A} be a structure with universe $A$, $\bar b\in A^s$, $1\le i\le k$, and $\psi(x_i,\bar y)$ a formula in the vocabulary of \str{A} with $\bar y=y_1\ldots y_s$. For 
\begin{eqnarray*}
\xi^{\bigvee}(\str{A}, \psi,\bar b):=\bigvee_{\substack{a\in A\\\str
A\models \psi(a,\bar b)}}X_{i,a}&\ \mbox{and}\ &
\xi^{\bigwedge}(\str{A}, \psi,\bar b):=\bigwedge_{\substack{a\in A\\\str
A\not\models \psi(a,\bar b)}}\neg X_{i,a}
\end{eqnarray*}
and for every functional assignment $S\se V$ with, say, $S(X_{i,a_0})=\textsc{True}$ we have
\[
S\text{ satisfies }\xi^{\bigvee}(\str{A}, \psi,\bar b)\iff\mathcal A\models\psi(a_0,\bar b)\iff S\text{ satisfies }\xi^{\bigwedge}(\str{A}, \psi,\bar b).
\]
\end{enumerate}
\end{lem}

\begin{prop}\label{pro:altd1}
Let $t\ge 2$.
\begin{enumerate}
\item If $t$ is even then $p\textup{-MC}(\Sigma_{t,1})\leq \textsc{WSat}(\Gamma^+_{t,1})$.
\item If $t$ is odd then $p\textup{-MC}(\Sigma_{t,1})\leq \textsc{WSat}(\Gamma^-_{t,1})$.
\end{enumerate}
\end{prop}
\proof 
Let $t\ge 2$ and $(\str A,\phi)$ an instance of
$p\textup{-MC}(\Sigma_{t,1})$. By the First-Order Normalisation Lemma we may
assume that $\phi\in\text{strict-}\Sigma_{t,1}$.
We shall define a propositional formula $\alpha$ of the desired syntactical
form  such that
\begin{equation}\label{eq:altd1-1}
\str A\models\phi\iff\alpha\text{ is $k$-satisfiable}.
\end{equation}
Suppose that
\[
\phi=\exists x_1\ldots\exists x_k\forall y_1\exists y_2\ldots
Q_{t-1}y_{t-1}\psi,
\]
where $Q_{t-1}=\exists$ if $t$ is odd and $Q_{t-1}=\forall$ if $t$ is even and $\psi$
is quantifier-free. We shall make further assumptions on $\phi$ when we branch
depending on $t$ later. Let $\bar y:=(y_1,\ldots,y_{t-1})$. Without loss of
generality we assume that $\psi$ is in negation normal form. We let $\Lambda$
denote the set of all literals occurring in $\psi$ (deviating from our earlier
proofs, where $\Lambda$ denoted a set of atoms). Recall that, because $\phi$ is in
strict-$\Sigma_{t,1}$, at most one of the
variables $x_1,\ldots,x_k$ occurs in a literal $\lambda\in\Lambda$.

The formula $\alpha$ will have propositional variables $X_{i,a}$ for all $a\in
A$ and $1\le i\le k$. The intended meaning of $X_{i,a}$ is: ``First-order
variable $x_i$ takes value $a$.'' Let $V$ be the set of all these
propositional variables.

Now assume that $t$ is even. Without loss of generality we may assume that 
$\psi=\bigwedge_{i=1}^\ell\bigvee_{j=1}^{m_i}\lambda_{ij}$ is in conjunctive
normal form, i.e.,
\[
\phi=\exists x_1\ldots\exists x_k\forall y_1\exists y_2\ldots
\forall y_{t-1}\bigwedge_{i=1}^\ell\bigvee_{j=1}^{m_i}\lambda_{ij},
\]
We use the formulas $\chi^+$ and $\xi^{\bigvee}(\ldots )$ of the preceding lemma and let
\[
\alpha:=\chi^+\wedge\bigwedge_{b_1\in A}\bigvee_{b_2\in
A}\ldots\bigwedge_{b_{t-1}\in
A}\bigwedge_{i=1}^\ell\bigvee_{j=1}^{m_i}\xi^{\bigvee}(\str{A},\lambda_{ij},b_1,\ldots,b_{t-1}).
\]
Clearly, $\alpha$ satisfies \eqref{eq:altd1-1} and can easily be transformed
into an equivalent $\Gamma_{t,1}^+$-formula.

For odd $t\ge 3$ we proceed similarly, except that we assume that $\psi$ is in disjunctive
normal form and that we replace $\chi^+$ and $\xi^{\bigvee}(\ldots )$ by 
$\chi^-$ and $\xi^{\bigwedge}(\ldots )$, respectively. 
\qed

\begin{prop}\label{pro:altd11}
\[
p\textup{-MC}(\Sigma_{1})\leq \textsc{WSat}(\Gamma^-_{1,2}).
\]
\end{prop}
It is straightforward to derive this proposition from the well known result
that $p\textup{-MC}(\Sigma_{1}[2])$ is reducible to the parameterized clique
problem (cf.~e.g.~\cite{gro01c}). However, to keep this paper self-contained we
give a direct proof.

\medskip\noindent
\textit{Proof of Proposition~\ref{pro:altd11}:}
Let $(\str A,\phi)$ be an instance of
$p\textup{-MC}(\Sigma_{1})$. By the First-Order Normalisation Lemma we may
assume that the vocabulary of $\phi$ is binary. We may further assume that
$\phi$ is of the form 
\[
\exists x_1\ldots\exists
x_k\bigvee_{p=1}^\ell\bigwedge_{q=1}^{m_p}\lambda_{pq},
\]
where each $\lambda_{pq}$ is a literal.

In a first step of the proof we shall define formulas
$\alpha_1,\ldots,\alpha_\ell\in\Gamma_{1,2}^-$ such that for $1\le p\le \ell$, 
\begin{equation}\label{eq:al-1}
\str A\models\exists x_1\ldots\exists
x_k\bigwedge_{q=1}^{m_p}\lambda_{pq}\iff\alpha_p\text{ is $k$-satisfiable}.
\end{equation}
Thus 
\begin{equation}\label{eq:al-2}
\str A\models\phi\iff\text{exists }p,1\le p\le\ell:\;
\alpha_p\text{ is $k$-satisfiable}.
\end{equation}
Let us fix $p$. We let $V_p$ be the set of propositional variables $X^p_{i,a}$
for $1\le i\le k$ and $a\in A$. Let $\chi^-_p$ be the corresponding formula
$\chi^-$ according to Lemma \ref{lem:mod}(1) for $V=V_p$.

 Similarly as $\chi^{\bigwedge}(\ldots )$ in Lemma \ref{lem:mod}(2), we  define, 
for $1\le q\le m_p$,
\[
\xi_{pq}:=\bigwedge_{\substack{a_1,a_2\in A\\\str
    A\not\models \lambda_{pq}(a_1,a_2)}}(\neg X^p_{i_1,a_1}\vee\neg
    X^p_{i_2,a_2}),
\]
where we assume that the free variables of $\lambda_{pq}$ are among
$x_{i_1},x_{i_2}$. Recall that the vocabulary of $\phi$ is binary, thus a
literal never has more than two free variables. 
For every functional
assignment $\{X^p_{1,a_1},\ldots,X^p_{k,a_k}\}\in 2^{V_p}$ we have
\[
\{X^p_{1,a_1},\ldots,X^p_{k,a_k}\}\text{ satisfies }\xi_{pq}\iff\mathcal
A\models\lambda_{pq}(a_{i_1},a_{i_2}).
\]
Thus
\[
\alpha_p:=\chi_p^-\wedge\bigwedge_{q=1}^{m_p}\xi_{pq}
\]
satisfies \eqref{eq:al-1}.

\medskip
By \eqref{eq:al-2}, it remains to define a formula $\alpha$ such that
\begin{equation}\label{eq:al-3}
\alpha\text{ is $k$-satisfiable}\iff\text{exists }p,1\le p\le\ell:\;
\alpha_p\text{ is $k$-satisfiable}.
\end{equation}
Let $V:=V_1\cup\ldots\cup V_{\ell}$. We call an assignment $S\in 2^V$
\emph{good} if there is a $p,1\le p\le\ell$ such that $S\subseteq V_p$. The
formula 
\[
\chi:=\bigwedge_{\substack{1\le i_1,i_2\le k\\a_1,a_2\in A\\1\le p_1< p_2\le\ell}}(\neg X^{p_1}_{i_1,a_1}\vee\neg
    X^{p_2}_{i_2,a_2})
\]
says that an assignment is good.  Note that if $S\subseteq V_p$ then $S$
satisfies $\alpha_{p'}$ for all $p'\neq p$, because variables only occur
negatively in $\alpha_{p'}$. Thus $S$ satisfies $\bigwedge_{r=1}^\ell\alpha_r$
if and only if $S$ satisfies $\alpha_p$. Therefore,
\[
\alpha:=\chi\wedge\bigwedge_{p=1}^\ell\alpha_p
\]
satisfies \eqref{eq:al-3}. Altogether, $(\str{A},\varphi)\mapsto (\alpha,k)$ is an \fpt-reduction.
\qed
\subsection{The W-hierarchy} We apply the results of the preceding two sections to the W-hierarchy. By definition the $t$th class of this hierarchy consists of all parameterized problems \fpt-reducible to the weighted satisfiability problem $\textsc{WSat}(\Omega_{t,d})$ for some $d$:

\begin{defi}For
 $t\ge 1$, $\W t:=[\{\textsc{WSat}(\Omega_{t,d})\mid d\ge t\}]^{\fpt}$.
\end{defi}
Putting all together, we get:
\begin{thm}\label{the:afo}For $t\geq 1$,
\[
\W t=[p\textup{-MC}(\Sigma_{t,1}[2])]^{\fpt}=[\{p\textup{-MC}(\Sigma_{t,u})\mid u\ge 1\}]^{\fpt}.
\] 
Moreover,
\begin{itemize}
\item if $t$ is even, $\W {t}=[\textsc{WSat}(\Gamma_{t,1}^{+})]^{\fpt}$;
\item if $t\geq 3$ is odd, $\W {t}=[\textsc{WSat}(\Gamma_{t,1}^-)]^{\fpt}$;
\item $\W {1}=[\textsc{WSat}(\Gamma_{1,2}^-)]^{\fpt}$.
\end{itemize}
\end{thm}
\proof All statements are immediate consequences of preceding results, e.g.:
\begin{eqnarray*}\textsc{WSat}(\Omega_{t,d})&\le &\textsc{WSat}(\Delta_{t,2^d}) \mbox{\quad (by Lemma \ref{lem:pn1})}\\
&\le&p\textup{-MC}(\Sigma_{t,1}[2])\mbox{\quad (by Proposition \ref{pro:afo} and Lemma \ref{lem:mcr})}\\
&\le&\textsc{WSat}(\Gamma_{t,2})\mbox{\quad (by Proposition \ref{pro:altd1} and Proposition \ref{pro:altd11})}.
\end{eqnarray*}
\qed
Now, Proposition~\ref{pro:afo} yields:

\begin{cor}\label{cor:col} For all $t,d\ge 1$,
\[
\textsc{WSat}(\Delta_{t+1,d})\in \W {t}.
\]
 \end{cor}
 
 If we identify $\W 0$ with \FPT, then the statement of the preceding
 corollary is true for $t=0$, too; in fact, there is even a polynomial time
 algorithm deciding for given $(\alpha,k)$ with $\alpha\in \Delta_{1,d}$ and
 $k\in\mathbb N$ if $\alpha $ is $k$-satisfiable.

The following corollary fills the gap that was left open in Remark~\ref{rem:graph}.

\begin{cor}\label{cor:gra}For all $t\geq 2,u\ge 1$,
\[
p\textup{-MC}(\Sigma_{t,u})\le p\textup{-MC}({\Sigma_{t,1}}[\textup{GRAPH}]).
\]
\end{cor}

\proof 
Let
$t\geq 2$, say $t=3$. By Theorem~\ref{the:afo}, it suffices to show
\[\textsc{WSat}(\Gamma_{t,1}^{-})\leq
p\textup{-MC}({\Sigma_{t,1}}[\textup{GRAPH}]).\] Fix $\alpha \in
\Gamma_{t,1}^{-}$ and $k\in\mathbb N$. Let $\str{G}$ be the graph obtained
from the tree of $\alpha$ by removing the leaves, identifying the nodes
corresponding to negative literals with the same variable and adding two
cycles of length 3 to its root $r$. We can assume that in \str{G} all branches
from the root to a leaf of \str{G} have the same length, namely 3. We say
that pairwise distinct $w_1,w_2,w_3, w_4$ with $Ew_1w_2,Ew_2w_3, Ew_{3}w_4$,
with $w_1=r$ and with $w_4=x$ ``witness that $x$ is a leaf''. Then, as formula
$\phi$ we can choose a $\Sigma_{3,1}$-formula equivalent to
\begin{align*}\exists x_1\ldots \exists x_k\exists x\exists u_1 \exists u_2\exists v_1\exists v_2 
\exists \bar w_1\ldots \exists \bar w_k (\mbox{``$x,u_1,u_2$ and $x,v_1,v_2$ are distinct cycles''}\hspace{2cm}\\
\wedge \bigwedge_{1\le i< j\le k} x_i\not=x_j\wedge \mbox{``$\bar w_1$ witness that $x_1$ is a leaf'' $\wedge\ldots \wedge $ ``$\bar w_k$ witness that $x_k$ is a leaf''}\hspace{.5cm}\\
\land \forall y((Exy\to \exists z(Eyz\land z\not= x\land \neg Ezx_1\land \ldots \land \neg Ezx_k))).
\end{align*}
Clearly, $(\alpha,k)\in \textsc{WSat}(\Gamma_{t,1}^{-})\iff  \str{G}\models \phi$.
\qed

Theorem \ref{the:afo} shows that for the weighted satisfiability problem for
$\Gamma_{t,d}$ the relevant class of formulas are the monotone ones in case
$t$ is even, and the antimonotone ones in case $t$ is odd. The
so-called
monotone and antimonotone collapse theorem due to Downey and Fellows
\cite{dowfel95b, dowfel95} states that for all $t,d\geq 1$,
\[
\textsc{WSat}(\Gamma_{2\cdot t,d}^{-})\in \W {2\cdot t-1} \qquad\mbox{and}\qquad \textsc{WSat}(\Gamma_{2\cdot t+1,d}^{+})\in \W {2\cdot t}.
\]
We get the following stronger result:

\begin{thm} For all $t,d\geq 1$,
\[
\textsc{WSat}(\Delta_{2\cdot t+1,d}^{-})\in \W {2\cdot t-1} \qquad\mbox{and}\qquad \textsc{WSat}(\Delta_{2\cdot t+2,d}^{+})\in \W {2\cdot t}.
\]
\end{thm}

\proof Fix $k\in \mathbb N$. First, consider a formula $\alpha$ in $\Gamma_{1,d}^{+}$ with variables $X_1,\ldots,
X_m$, say
\[
\alpha=\bigwedge_{i\in I}( Y_{i1}\lor\ldots \lor Y_{ir_i}).
\]
Compute the minimal covers of size $\leq k$ of the hypergraph $\str{H}=(H,E)$,
where $H:=\{1,\ldots, m \}$ and $$E:=\{\{i_1,\ldots, i_{r_i} \} \mid \mbox{for
  some $i\in I$: $X_{i_1}=Y_{i1}$ ,\ldots, $X_{i_{r_i}}=Y_{ir_i}$}\}.$$
For
every such cover $C$ let $\gamma_C$ be the conjunction of the variables $X_j$
with $j\in C $. Then, $\gamma_C$ is the conjunction of at most $k$ variables.
With respect to assignments of $\alpha$ of weight $k$, the formulas $\alpha$
and $\bigvee_{C \textup{\ cover}}\gamma_C$ are equivalent.

Now, let $\beta\in \Delta_{2\cdot t+2,d}^{+}$ and $k\in\mathbb N$. We replace every subformula $\alpha\in\Gamma_{1,d}^{+}$ by the corresponding $\bigvee_{C \textup{\ cover}}\gamma_C$, thus obtaining a formula $\beta^*$ in $\Delta_{2\cdot t+1,k}^+$. Then, the result follows from Corollary \ref{cor:gdk}. (In case $\beta^*$ but not $\beta$ has $e<k$ variables, we check if $\beta^*$ is $e$-satisfiable.)

The proof for $\textsc{WSat}(\Delta_{2\cdot t+1,d}^{-})$ is obtained by treating subformulas in $\Delta_{1,d}^-$ in the dual way.  
\qed

\section{Back and forth between propositional and first-order logic: 
  the extensions}\label{sec:ba2}

\subsection{The $\textup{W}^*$-hierarchy}\label{ss:star}
In \cite{dowfeltay96}, Downey, Fellows, and Taylor introduced the
$\textup{W}^*$-hierarchy and showed that the first two levels of the
$\textup{W}^*$-hierarchy coincide with first the two levels of the
$\textup{W}$-hierarchy (\cite{dowfeltay96}, \cite{dowfel98}).  We first recall
the definition of the $\textup{W}^*$-hierarchy and then give complete
model-checking problems for the classes of this hierarchy. This
characterization allows simple proofs of $\textup{W}^*[1]=\textup{W}[1]$ and
$\textup{W}^*[2]=\textup{W}[2]$.
 
The crucial difference between the W-hierarchy and the
$\textup{W}^*$-hierarchy is that instead of being fixed, in the definition of
the $\textup{W}^*$-hierarchy the depth is treated as a parameter.

For a set $\Gamma$ of propositional formulas we let

\npprob{ \textsc{WSat$^*(\Gamma)$}}{$k\in\mathbb N$ and $\alpha\in\Gamma$ such
  that the depth of $\alpha$ is at most $k$}{$k$.}{Decide if $\alpha$ is $k$-satisfiable}

For every $t\ge 0$ we let $\Omega_t$ denote the set of all propositional
formulas of weft at most $t$.

\begin{defi}
For $t\ge 1$,
\[
\textup{W}^*[t]:=[\textsc{WSat}^*(\Omega_{t})]^{\fpt}.
\]
\end{defi} 

Before we turn to the first-order characterisation of the 
$\textup{W}^*$-hierarchy, we normalise the propositional formulas involved.
For $k\ge 1$ we define two new families $\Gamma^*_{t,k}$ and
$\Delta^*_{t,k}$ of propositional formulas. We use $\wedge_{i=1}^k\alpha_{i}$ as an
abbreviation for the formula
$(\cdots((\alpha_1\wedge\alpha_2)\wedge\alpha_3)\cdots\wedge\alpha_k)$.
Similarly, we use $\vee_{i=1}^k\alpha_{i}$.
\begin{itemize}
\item
We let
$\Gamma^*_{1,k}=\Gamma_{1,k}$ and $\Delta^*_{1,k}=\Delta_{1,k}$.
\item
For $t\ge 2$, we let $\Gamma^*_{t,k}$ be the class of all formulas of the form
\[
\bigwedge_{i\in I}\vee_{j=1}^k\alpha_{ij}
\]
where $I$ is an arbitrary (finite) index set and $\alpha_{ij}\in
\Gamma^*_{t-1,k}\cup\Delta^*_{t-1,k}$ for all $i\in I,1\le j\le k$. Similarly,
we let $\Delta^*_{t,k}$ be the class of all formulas of the form
\[
\bigvee_{i\in I}\wedge_{j=1}^k\alpha_{ij}
\]
where $I$ and the $\alpha_{ij}$ are as above.
\end{itemize}

Observe that $\Gamma^*_{t,k}\cup\Delta^*_{t,k}\subseteq\Omega_{t,t\cdot k}$. 

The following lemma, which may be viewed as the starred analogon of
Lemma~\ref{lem:pn1}, is essentially due to Downey, Fellows, and
Taylor~\cite{dowfeltay96}.
Denote by \textsc{Prop} the class of all propositional formulas.
\begin{lem}\label{lem:pnstar}
  Let $t\ge 1$. Then there is an fpt-algorithm that assigns to every instance $(\alpha,k)$ of $\textsc{WSat}^*(\Omega_{t})$ an instance $(\beta,\ell) $ of $\textsc{WSat}^*(\textsc{Prop})$ with $\beta\in \Delta^*_{t+1,\ell}$.
\end{lem}

\proof
By induction on $t\ge 1$ we first prove that every formula $\alpha$ in $\Omega_{t,k}$ whose
outermost connective is a big conjunction is equivalent to a formula  in
$\Gamma^*_{t,2^k}$ and simultaneously that 
every formula in $\Omega_{t,k}$ whose
outermost connective is a big disjunction is equivalent to a formula  in
$\Delta^*_{t,2^k}$. 

Suppose that $t\ge 1$ and $\alpha\in\Omega_{t,k}$ is of the form
$\bigwedge_{i\in I}\beta_i$. By the induction hypothesis, we can assume that
each $\beta_i$ is a Boolean combination of at most $2^k$ formulas in
$\Gamma^*_{t-1,2^k}\cup\Delta^*_{t-1,2^k}$ or, if $t=1$, propositional variables.
Transforming these Boolean combinations into conjunctive normal form, which
can be achieved by an \fpt-reduction since the number (at most $2^k$) of formulas is bounded
in terms of  the parameter, and merging the outermost conjunctions we obtain a formula  
of the desired form.  Formulas $\alpha$ whose outermost connective is a big
disjunction can be treated analogously.

Now it easily follows that there is an fpt-algorithm that assigns to every instance $(\alpha,k)$ of $\textsc{WSat}^*(\Omega_{t})$ a formula  $\alpha'\in \Delta^*_{t+1,2^k}$ such that ($\alpha$ is $k$-satisfiable $\iff$ $\alpha'$ is $k$-satisfiable). Let $\alpha'=\bigvee_{i\in I}\wedge_{j=1}^{2^k}\alpha'_{ij}$ and let $X_1,\ldots, X_{2^k+1-k}$ be new propositional variables. We set $\alpha'_{i\ 2^k+1}:=\bigwedge_{m=1}^{2^k+1-k}X_m$, for $i\in I$, and 
\[
\beta=\bigvee_{i\in I}\wedge_{j=1}^{2^k+1}\alpha'_{ij}.
\]
Then, $\beta\in \Delta^*_{t+1,2^k+1}$ and ($\alpha'$ is $k$-satisfiable $\iff$ $\beta$ is $2^k+1$-satisfiable). Therefore, $(\alpha,k)\mapsto (\beta,2^k+1)$ is the desired reduction.
\qed

We turn to the characterisation of $\textup{W}^*[t]$ in terms of a complete
model-checking problem.  To get the corresponding fragment of first-order
logic, we first point out a closure property of the classes $\Sigma_t$ not
shared by the $\Sigma_{t,u}$.  The closure of $\Sigma_{t,u}$ under this
operation yields the desired fragment.

The formula
\begin{equation}\label{equ:si3}
\exists \bar x(\forall  y_1\forall y_2\exists  z_1\exists z_2 \psi \wedge\exists v_1 \exists v_2\forall w_1\forall w_2 \chi)
\end{equation}
with quantifier-free $\psi(\bar x,\bar y,\bar z)$ and $\chi(\bar x,\bar v,\bar w)$ is an existential quantification of a
Boolean combination of $\Sigma_2$-formulas; it is logically equivalent to the
$\Sigma_3$-formula
\begin{equation*}\exists \bar x\exists v_1\exists v_2\ \forall y_1\forall y_2\ \forall w_1\forall w_2\ \exists z_1\exists z_2(\psi \wedge\chi);
\end{equation*}
more generally, every existential quantification of a Boolean combination of
$\Sigma_2$-formulas is equivalent to a $\Sigma_3$-formula.

The class $\Sigma_{3,2}$ does not have this closure property, the formula in
(\ref{equ:si3}) is an existential quantification of a Boolean combination of
$\Sigma_{2,2}$-formulas with {\em all} blocks of length $\le 2$, but, in
general, it is not logically equivalent to a formula in $\Sigma_{3,2}$. The
class $\Sigma^*_{3,2}$ and (the classes $\Sigma^*_{t,u}$) are defined in such
a way that they have this closure property.

For this purpose, first define the set $\Theta_{t,u}$ of first-order formulas
by induction:
\[
\begin{array}{rcl}
\Theta_{0,u}&:=&\mbox{the set of quantifier-free formulas}\\
\Theta_{t+1,u}&:=&\mbox{Boolean combinations of formulas of the form $\exists y_1\ldots \exists y_u \psi$ with $\psi\in \Theta_{t,u}$}
\end{array}
\]
Now let $\Sigma^*_{t,u}$ be the set of formulas of the form
\[
\exists x_1\ldots \exists x_k \psi
\]
where $\psi\in\Theta_{t-1,u} $.

As for the un-starred version, a $\Sigma^*_{t,u}$-formula is in
$\textup{strict-}\Sigma^*_{t,u}$ if each atomic subformula contains at most one
variable of the first block of its prefix.  
We leave it to the reader to
verify the following lemma, which is the analogon for $\Sigma^*_{t,u}$ of part
(1) of the First-Order Normalisation Lemma.

\begin{lem}\label{lem:wst}
  For $t\ge 2,u\geq 1$, $p\textup{-MC}(\Sigma^*_{t,u})\leq
  p\textup{-MC}(\textup{strict-}\Sigma^*_{t,1}[2]).$
\end{lem}

The following lemma is a stronger version of Lemma~\ref{lem:g1d}:

\begin{lem}\label{lem:g1dstar} For all $k\ge 1$ and for all
  formulas $\alpha:=\wedge_{i=1}^k\alpha_i$, where
  $\alpha_1,\ldots,\alpha_k\in\Gamma_{1,k}\cup\Delta_{1,k}$, there are
\begin{itemize}
\item
a structure $\str A:=\str A_{\wedge,\alpha,k}$ with universe $A:=\{1,\ldots, m \}$, where the  variables of $\alpha$ are among
$X_1,\ldots,X_m$, 
\item a quantifier-free formula $\psi:=\psi_{\wedge,k}$ depending
  only on $k$
\end{itemize}
such that the mapping $(\alpha,k)\mapsto(\str A,\psi)$ is
fixed-parameter tractable and for pairwise distinct $m_1,\ldots,m_k\in A$,
\[
\{X_{m_1},\ldots, X_{m_k}\}\text{ satisfies }\alpha\iff\str{A}\models \psi(m_1,\ldots, m_k).
\]
\end{lem}

\proof For $1\le i\le k$, if $\alpha_i\in\Gamma_{1,k}$ we let $\str A_i:=\str
A_{\bigwedge,\alpha_i(X_1,\ldots, X_m),k,k}$ and $\psi_i:=\psi_{\bigwedge,k,k}$ be the
structure and sentence obtained from Lemma~\ref{lem:g1d}, and if
$\alpha_i\in\Delta_{1,k}$ we let $\str A_i:=\str A_{\bigvee,\alpha_i(X_1,\ldots, X_m),k,k}$ and
$\psi_i:=\psi_{\bigvee,k,k}$ be the structure and sentence obtained from
Corollary~\ref{cor:dis}. We let $\tau_i$ be the vocabulary obtained from the
vocabulary of $\mathcal A_i$ and $\psi_i$ by replacing each relation symbol
$R$ by a new symbol $R_i$ of the same arity. We let $\mathcal A_i'$ and
$\psi_i'$ be the $\tau_i$-structure and sentence obtained from $\mathcal A_i$
and $\psi_i$, respectively, by replacing each relation symbol $R$ by $R_i$.

Note that the universe of $\mathcal A_1',\ldots\mathcal A_k'$ is
$\{1,\ldots,m\}=:A$. Let $\tau:=\bigcup_{i=1}^k\tau_i$, and let $\mathcal A'$
be the $\tau$-structure with universe $A$ and $R_i^{\mathcal
  A'}:=R_i^{\mathcal A_i'}$ for all relation symbols $R_i\in\tau_i$ and $1\le
i\le k$.

Note that for $1\le i\le k$ and pairwise distinct $m_1,\ldots, m_k\in A$
\[
\mathcal A\models \psi_i'(m_1,\ldots, m_k)\iff\mathcal A_i\models\psi_i(m_1,\ldots, m_k).
\]
Thus by Lemma~\ref{lem:g1d} and  Corollary~\ref{cor:dis}, for pairwise
distinct $m_1,\ldots,m_k\in A$ we have
  \begin{eqnarray*}
    X_{m_1},\ldots, X_{m_k} \mbox{\ satisfies $\alpha$}&\iff& 
    \str{A}\models \bigwedge_{i=1}^k\psi_i'(m_1,\ldots, m_k).
  \end{eqnarray*}
  The only remaining problem is that the formula
  $\bigwedge_{i=1}^k\psi_i'(x_1,\ldots, x_k)$ depends on $\alpha$. But
  actually it only depends on which of $\alpha_1,\ldots,\alpha_k$ are in
  $\Gamma_{1,k}$ and which in $\Delta_{1,k}$. We introduce $k$ new unary
  relation symbols $C_1,\ldots,C_k$ and let $\mathcal A$ be the expansion of
  $\mathcal A'$ with
\[
C_i^{\mathcal A}:=
\begin{cases}
A&\text{if }\alpha_i\in\Gamma_{1,k},\\
\emptyset&\text{if }\alpha_i\in\Delta_{1,k}.
\end{cases}
\]
We let $\psi_{\bigwedge}^i$ be the formula obtained from the formula
$\psi_{\bigwedge,k,k}$ of Lemma~\ref{lem:g1d} by replacing each relation symbol $R$
by the corresponding $R_i$ and define $\psi_{\bigvee}^i$ accordingly. Thus
$\psi_i$ is either $\psi_{\bigwedge}^i$ or $\psi_{\bigvee}^i$, depending on
whether $\alpha_i\in\Gamma_{1,k}$ or $\alpha_i\in\Delta_{1,k}$.
Finally, we let
\[
\psi(x_1,\ldots,x_k):=\bigwedge_{i=1}^k\Big(\big(C_ix_1\to \psi_{\bigwedge}^i(x_1,\ldots,x_k)\big)
\wedge\big(\neg C_ix_1\to \psi_{\bigvee}^i(x_1,\ldots,x_k)\big)\Big).
\]
\qed

\begin{cor}\label{cor:disstar}\sloppy
  For all $k\ge 1$ and for all
  formulas $\alpha:=\vee_{i=1}^k\alpha_i$, where
  $\alpha_1,\ldots,\alpha_k\in\Gamma_{1,k}\cup\Delta_{1,k}$, there are
\begin{itemize}
\item
a structure $\str A$ with universe $A:=\{1,\ldots, m \}$, where the    variables of $\alpha$ are among
$X_1,\ldots,X_m$, 
\item a quantifier-free formula $\psi_{\vee,k}$ depending
  only on $k$
\end{itemize}
such that the mapping $(\alpha,k)\mapsto(\str A,\psi_{\vee,k})$ is
fixed-parameter tractable and
for  pairwise distinct $m_1,\ldots,m_k\in A$
\[
\{X_{m_1},\ldots, X_{m_k}\}\text{ satisfies }\alpha\iff\str{A}\models \psi_{\vee,k}(m_1,\ldots, m_k).
\]
\end{cor}

The following two propositions will yield the characterisation of the
W$^*$-hierarchy in terms of model-checking problems.

\begin{prop}\label{pro:afo*} For  $t\geq 1$,  
\[
\textsc{WSat}^*(\Omega_{t})\leq p\textup{-MC}(\Sigma^*_{t,2}).
\] 
\end{prop}

\proof  Recall the proof of Proposition~\ref{pro:afo}; we proceed very
similarly here and mainly point out where the the proofs differ.
Fix $t\ge 1$. 
Let $(\alpha,k)$ be an instance of 
$\textsc{WSat}^*(\Omega_{t})$. 
We shall construct a structure $\mathcal
A$ and a $\Sigma_{t,2}$-sentence $\phi$ such that
\begin{equation}\label{equa:wstar}
\alpha\text{ is $k$-satisfiable}\iff \mathcal A\models \phi. 
\end{equation}

By Lemma~\ref{lem:pnstar}, we may assume that
$\alpha\in\Delta^*_{t+1,k}$. Let the variables of $\alpha$ be
$X_1,\ldots,X_m$. As in the proof of Proposition~\ref{pro:afo}, the structure
$\str A$ consists of two parts: a tree  representing the parse tree
of the formula $\alpha$ and, attached to the leaves of the tree, a structure on the
variables representing the innermost subformulas.

However, a formula in $\Delta^*_{t+1,k}$ is not as regular as a formula in
$\Delta_{t+1,d}$, and therefore the definition of the tree is more involved.
In particular, some of the nodes and edges of the tree carry additional information. 

First, we let \str{T} be the tree
 obtained from the ``parse tree'' of $\alpha$ by
removing all nodes that correspond to  subformulas of $\alpha$ in $\Gamma_{1,k}\cup\Delta_{1,k}$. Thus, the leaves correspond to subformulas of the  form $\vee_{i=1}^k\beta_i$ or
  $\wedge_{i=1}^k\beta_i$, where $\beta_1,\ldots,\beta_k\in\Gamma_{1,k}\cup\Delta_{1,k}$. In addition to the relation symbol $E$ for the edge relation of this tree (directed from the root to the leaves), we have  binary
relation symbols $\{E_1,\ldots,E_k\}$ and unary relation symbols
$K_1,\ldots, K_k$ and $\textup{Root}$ whose interpretation in \str{T} is fixed by the following clauses: Let $u$ be a node of \str{T} and $\beta$ the subformula of $\alpha$ corresponding to the node $u$.
\begin{itemize}
\item If $u$ is the root of the tree, then $\textup{Root}^{\str{T}}u$;
\item If $\beta=\vee_{i=1}^k\beta_i$ or $\beta=\wedge_{i=1}^k\beta_i$, where
  $\beta_1,\ldots,\beta_k\in\Gamma^*_{s,k}\cup\Delta^*_{s,k}$ for some $s\ge
  2$, then, for $1\le j\le k$, $E^{\str{T}}_juu_j$ where $u_j$ is the child of $u$ corresponding to $\beta_j$. Moreover, $K^{\str{T}}_ju$ if $\beta_j\in \Gamma^*_{s,k}$.
\end{itemize}
Note that we encode the information on whether a subformula
$\beta\in\Gamma^*_{s,k}\cup\Delta^*_{s,k}$ is in $\Gamma_{s,k}^*$ or in
$\Delta_{s,k}^*$ by putting the parent  into the
corresponding relation $K_j$ if $\beta$ is in $\Gamma_{s,k}^*$. The reason
that we choose such a counter-intuitive encoding is that we need the
information about the child at the parent in order to pick the right
quantifier to access the child. The definition of the formulas
$\psi^{s+1}_{\bigwedge}$ and $\psi^{s+1}_{\bigvee}$ below will clarify this.

 The second part of the structure $\mathcal A$ we are heading for is defined as in the proof
of Proposition~\ref{pro:afo}, except that now the leaves of the tree are
formulas of the form $\vee_{i=1}^k\beta_i$ or $\wedge_{i=1}^k\beta_i$, where
$\beta_1,\ldots,\beta_k\in\Gamma_{1,k}\cup\Delta_{1,k}$, and we have to use
Lemma~\ref{lem:g1dstar} and Corollary~\ref{cor:disstar} instead of
Lemma~\ref{lem:g1d} and Corollary~\ref{cor:dis}.

 We define formulas
$\psi^s_{\wedge}(y,x_1,\ldots,x_k)$ and 
$\psi^s_{\vee}(y,x_1,\ldots,x_k)$ for $1\le s\le t+1$, and formulas 
$\psi^s_{\bigwedge}(y,x_1,\ldots,x_k)$ and
$\psi^s_{\bigvee}(y,x_1,\ldots,x_k)$  for $2\le s\le t+1$ such that for every node $u\in T$ corresponding to a subformula $\beta$
 and for all
$a_1,\ldots,a_k\in\{1,\ldots,m\}$ we have
\begin{enumerate}
  \RNC{\labelenumi}{(\roman{enumi})}
\item
If $\beta=\wedge_{i=1}^k\beta_i$, where
$\beta_1,\ldots,\beta_k\in\Gamma_{s,k}\cup\Delta_{s,k}$, then 
\[
\{X_{a_1},\ldots,X_{a_k}\}\text{ satisfies }\beta\iff\str
A\models\psi^s_{\wedge}(u,a_1,\ldots,a_k).
\]
\item
If $\beta=\vee_{i=1}^k\beta_i$, where
$\beta_1,\ldots,\beta_k\in\Gamma_{s,k}\cup\Delta_{s,k}$, then 
\[
\{X_{a_1},\ldots,X_{a_k}\}\text{ satisfies }\beta\iff\str
A\models\psi^s_{\vee}(u,a_1,\ldots,a_k).
\]
\item
If $\beta\in\Gamma_{s,k}^*$, then
$
\big(\{X_{a_1},\ldots,X_{a_k}\}\text{ satisfies }\beta\iff\str
A\models\psi^s_{\bigwedge}(u,a_1,\ldots,a_k)\big).
$
\item
If $\beta\in\Delta_{s,k}^*$, then
$
\big(\{X_{a_1},\ldots,X_{a_k}\}\text{ satisfies }\beta\iff\str
A\models\psi^s_{\bigvee}(u,a_1,\ldots,a_k)\big).
$
\end{enumerate}
We let $\psi^1_{\wedge}(y,x_1,\ldots,x_k)$ be the formula obtained from the
formula $\psi_{\wedge,k}(x_1,\ldots,x_k)$ of Lemma~\ref{lem:g1dstar} by
replacing each atomic subformula $Rx_1\ldots x_r$ by $R'yx_1\ldots x_r$ (compare this to  the proof of Proposition~\ref{pro:afo}).
Similarly, we define $\psi^1_{\vee,k}(y,x_1,\ldots,x_k)$ using the formula
$\psi_{\vee}(x_1,\ldots,x_k)$ of Corollary~\ref{cor:disstar}.

For $s\ge 1$, we let 
\begin{align*}
\psi^{s+1}_{\bigwedge}(y,x_1,\ldots,x_k)&:=\forall
z\big(Eyz\to\psi^{s}_\vee(z,x_1,\ldots,x_k)\big),\\
\psi^{s+1}_{\bigvee}(y,x_1,\ldots,x_k)&:=\exists
z\big(Eyz\wedge\psi^{s}_\wedge(z,x_1,\ldots,x_k)\big),\\
\psi^{s+1}_\wedge(y,x_1,\ldots,x_k)&:=\bigwedge_{i=1}^k
\Big(\begin{array}[t]{@{}c@{\;}l}
&\big(K_iy\to\forall
z\big(E_iyz\to\psi^{s+1}_{\bigwedge}(z,x_1,\ldots,x_k)\big)\big)\\
\wedge&
\big(\neg K_iy\to\exists
z\big(E_iyz\wedge\psi^{s+1}_{\bigvee}(z,x_1,\ldots,x_k)\big)\big)
\Big),
\end{array}\\
\psi^{s+1}_\vee(y,x_1,\ldots,x_k)&:=\bigvee_{i=1}^k
\Big(\begin{array}[t]{@{}c@{\;}l}
&\big(K_iy\wedge\forall
z\big(E_iyz\to\psi^{s+1}_{\bigwedge}(z,x_1,\ldots,x_k)\big)\big)\\
\vee&
\big(\neg K_iy\wedge\exists
z\big(E_iyz\wedge\psi^{s+1}_{\bigvee}(z,x_1,\ldots,x_k)\big)\big)
\Big).
\end{array}
\end{align*}
It is easy to see now that  these formulas
satisfy (i)--(iv). Furthermore, 
$\psi^1_{\wedge}$ and $\psi^1_{\vee}$ are quantifier-free and,  by a simultaneous induction on 
 $s\ge 1$,
\begin{itemize}
\item $\psi^{s+1}_{\bigwedge}$ can be transformed into a formula of the form
$\forall y\chi$, where $\chi\in\Theta_{s-1,2}$;
\item $\psi^{s+1}_{\bigvee}$ can be transformed into a formula of the form
$\exists z\chi$, where $\chi\in\Theta_{s-1,2}$;
\item $\psi^{s+1}_\wedge$ and $\psi^{s+1}_\vee$
can easily be transformed into a formula in $\Theta_{s,2}$.
\end{itemize}
We let 
\[
\phi:=\exists x_1\ldots\exists x_k\exists y(\bigwedge_{1\le i<j\le k}x_i\not= x_j\wedge \textup{Root}\, y\wedge\psi^{t+1}_{\bigvee}(y,x_1,\ldots,x_k)).
\]
It is easy to see that $\phi$ is equivalent to a formula in $\Sigma^*_{t,2}$.
\qed

\begin{prop} 
  For all $t, u\geq 1$, $p\textup{-MC}(\Sigma^*_{t,u})\leq
  \textsc{WSat}^*(\Omega_{t})$.
\end{prop}

\proof The proof essentially duplicates the arguments of the proof of
Proposition \ref{pro:altd1}. The additional disjunctions and conjunctions
between blocks of quantifiers in a $\Sigma^*_{t,u}$-formula $\phi$ yield
additional connectives in the propositional formula we look for.
\qed 

By Lemma \ref{lem:wst} and the preceding propositions we get:

\begin{thm}\label{the:hsw*}
  For $t,u\geq 1$,
  \[
  \textup{W}^*[t]=
  [p\textup{-MC}(\Sigma^*_{t,u})]^{\fpt}
  =[p\textup{-MC}(\Sigma^*_{t,1}[2])]^{\fpt}.
  \]
\end{thm}

\begin{cor}$\textup{W}^*[1]=\W 1$.
\end{cor}
\proof Since $\Sigma^*_{1,u}=\Sigma_{1,u}=\Sigma_1$, this is immediate by Theorem \ref{the:afo} and Theorem \ref{the:hsw*}. 
\begin{cor}$\textup{W}^*[2]=\W 2$.
\end{cor}
\proof Again by Theorem \ref{the:afo} and Theorem \ref{the:hsw*}, it suffices to show that 
\[
p\textup{-MC}(\Sigma^*_{2,u})\leq p\textup{-MC}(\Sigma_{2,u}).
\]
So let $\str A$ be a structure and $\phi$ a $\Sigma^*_{2,u}$-sentence. We can
assume that $\phi$ has the form
\[
\exists x_1\ldots \exists x_\ell \bigvee_{i\in I}\bigwedge_{j\in J_i}\psi_{ij},
\]
where $I$ and  the $J_i$ are finite sets and the $\psi_{ij}$ are formulas in $\Sigma_1\cup\Pi_1$ with quantifier block of length $\leq u$. First we replace  the disjunction $\bigvee_{i\in I}$ in $\phi$ by an existential quantifier. For this purpose, we add to the vocabulary $\tau$ of \str{A} unary relation symbols $R_i$ for $i\in I$ and consider an expansion $(\str{A},(R_i^A)_{i\in I})$ of \str{A}, where $(R_i^A)_{i\in I}$ is  a partition of $A$ into nonempty disjoint sets. Then
\begin{eqnarray*}\str{A}\models\phi&\iff&(\str{A},(R_i^A)_{i\in I})\models \exists x_1\ldots \exists x_\ell \exists y\bigwedge_{j\in J_i}(\neg R_iy\vee\psi_{ij}).
\end{eqnarray*}
Altogether, we can assume that $\phi$ has the form
\[
\exists x_1\ldots \exists x_\ell\bigwedge_{j=1}^m\psi_{j},
\]
where for some quantifier-free $\chi_j$
\[
\psi_{j}=\exists \bar y_j\chi_j \mbox{\quad for $j=1,\ldots, s$}
\]and
\[
\psi_{j}=\forall \bar z\chi_j \mbox{\quad for $j=s+1,\ldots, m$}.
\]
Here, $\bar y_1,\ldots ,\bar y_s,\bar z$ are sequences of length $\leq u$ and
 we can assume that any two of them have no variable in common.  But then $\phi$ is equivalent to the  $\Sigma_{2,u}$-formula:
\[
\exists x_1\ldots \exists x_\ell\exists\bar y_1\ldots \exists \bar y_s\forall \bar z \bigwedge_{j=1}^m\chi_j .
\]
\qed 

Unfortunately, the argument of the preceding proof cannot be extended to an
inductive proof of $\textup{W}^*[t]=\W t$ for all $t\ge 2$. To see this,
observe that for an instance
$(\str{A},\varphi)$ of $p\textup{-MC}(\Sigma^*_{3,u})$, in the same way we
would obtain an equivalent formula
\[
\varphi':=\exists x_1\ldots \exists x_\ell\exists\bar y_1\ldots \exists \bar y_s\forall \bar z \bigwedge_{j=1}^m\chi_j,
\]
where now the $\chi_j$ are Boolean combinations of
$\Sigma^*_{2,u}\cup\Pi^*_{2,u} $-formulas with all quantifier blocks of length
at most $u$. But now the existential quantifiers in the $\chi_j$ cannot be
transferred to the leading existential block in $\varphi'$, they are blocked
by the universal quantifiers.

\subsection{The A-hierarchy}\label{sec:alt}
Originally, the A-hierarchy was defined by means of halting problems:
$\textup{A}[\ell]$ (where $\ell\in\mathbb N$) has as complete problem the
halting problem for alternating Turing machines with $\ell-1$ alternations
(and existential starting state), parameterized by the number of steps. In
\cite{flugro01}, it was shown that $
\textup{A}[\ell]=[\{p\textup{-MC}(\Sigma_{\ell}[r])\mid r\ge 1\}]^{\fpt} $. In
view of part 2 of the Normalisation Lemma this yields
$$
\textup{A}[\ell]=[p\textup{-MC}(\Sigma_{\ell})]^{\fpt},
$$
which, in this paper, we take as definition of the A-hierarchy. Since
$\Sigma_{1,u}=\Sigma_1$ and $\Sigma_{\ell,u}\subseteq\Sigma_{\ell}$, we have
\[
\textup{W}[1]=\textup{A}[1]\qquad\mbox{and for $\ell\ge 2$:}\ \ 
\textup{W}[\ell]\subseteq\textup{A}[\ell].
\]
In this section we derive a characterisation of the A-hierarchy in terms of
weighted satisfiability problems for classes of propositional formulas.

We saw in the preceding sections that a single universal quantifier (or
equivalently, a block of bounded length of universal quantifiers) in a
first-order formula translates into a $\bigwedge$ in the corresponding
propositional formula, and similarly, an existential quantifier translates
into a $\bigvee$. As the proof of Proposition \ref{pro:altd1} shows the
leading (unbounded) block $\exists x_1\ldots \exists x_k$ yields, on the side
of propositional logic, the weight or parameter $k$ and the propositional
variables $X_{i,a}$ (with $1\le i\le k$ and with $a$ ranging over the universe
of the given structure). Since in $\textup{A}[\ell]$ we have $\ell$
alternating (unbounded) blocks, we have to consider alternating weighted
satisfiability problems for classes of propositional formulas. Such problems
were already introduced by Abrahamson, Downey, and Fellows in
\cite{abrdowfel95} when they considered quantified boolean (propositional) logic.

Let $\Gamma$ be a set of propositional formulas (as defined in Section
\ref{sec:pre}) and $\ell\geq 1$. The \emph{$\ell$-alternating weighted
  satisfiability problem \textsc{AWSat$_{\ell}(\Gamma)$} for formulas in
  $\Gamma$} is the following problem:
\npprob{\textsc{AWSat$_{\ell}(\Gamma)$}}{$\alpha\in\Gamma$ and a partition
  $I_1\, \dot{\cup}\, \ldots \, \dot{\cup}\, I_{\ell}$ of the propositional
  variables of $\alpha$}{$k_1,\ldots ,k_{\ell}\in\mathbb N$.}{Decide if there
  is a size $k_1$ subset $S_1$ of $I_1$ such that for every size $k_2$ subset
  $S_2$ of $I_2$ there exists \ldots such that the truth value assignment
   $S_1\cup\ldots \cup S_{\ell}$ satisfies $\alpha$}
Thus, $\textsc{AWSat}_{1}(\Gamma)=\textsc{WSat}(\Gamma)$. Generalising the
definition
\[
 \W t:=[\{\textsc{WSat}(\Omega_{t,d}) \mid d\geq t\}]^{\fpt}
\]
of the classes of the W-hierarchy on the alternating level, we define the
parameterized complexity class $\textup{A}[\ell,t]^{\fpt}$ by
\[
\textup{A}[\ell,t]:=
[\{\textsc{AWSat$_{\ell}(\Omega_{t,d})$} \mid d\geq t\}]^{\fpt}.
\]
Thus, $\textup{W}[t]=\textup{A}[1,t]$ and as the main result of this section
will show, $\textup{A}[\ell]=\textup{A}[\ell,1]$, which yields the desired
characterisation of the A-hierarchy in terms of propositional logic. Thus, the
family of classes $\textup{A}[\ell,t]$, which we may call the
\emph{A-matrix}, contains the classes of the W-hierarchy
and the classes of the A-hierarchy.

We turn to a model-checking characterisation of this family: The propositional
formulas in the defining problem of $\textup{A}[\ell,t]$ contain $\ell$
``weighted alternations'' and at most $t$ (nested) big conjunctions or big
disjunctions. As we remarked above, the $\ell$ weighted alternations translate
into $\ell$ alternating blocks of quantifiers and the $t$ (nested) big
conjunctions or big disjunctions into $t$ further quantifiers; the first of
them can be merged with the last alternating block, so we expect that
\[
\textup{A}[\ell,t]=[p\textup{-MC}(\Sigma^{\ell,t-1})]^{\fpt},
\]
where for $\ell \geq 1$ and $m\geq 0$ we denote by $\Sigma^{\ell,m}$ the class
of first-order formulas of the form
\[
\exists \bar x_1\forall \bar x_2\ldots Q_{\ell}\bar
x_{\ell}Q_{\ell+1}x_{\ell+1}\ldots Q_{\ell+m}x_{\ell+m}\psi
\]
where $\psi$ is quantifier-free, all $Q_i\in \{\exists, \forall \}$, and
$Q_i\not= Q_{i+1}$. Note that $\bar x_{\ldots}$ denotes a finite sequence of
variables, thus the formula starts with $\ell$ unbounded blocks of
quantifiers. Hence,
\begin{itemize}
\item $\Sigma^{\ell,0}=\Sigma_{\ell}$.
\item For $t\geq 1$, $\Sigma^{1,t-1}=\Sigma_{t,1}$.
\end{itemize}
It should be clear how the class $\Pi^{\ell,m}$ of formulas is defined.

We call a $\Sigma^{\ell,m}$-formula \emph{strict} if each atomic subformula 
contains at most one variable from the first $\ell$, the unbounded blocks of
quantifiers. 
Again, part 1 of the First-Order Normalisation Lemma generalizes
(with essentially the same proof) to $\Sigma^{\ell,m}$. We state the result
and leave its verification to the reader:

\begin{lem}\label{lem:nla}For   $\ell,m\geq 1$, 
  $p\textup{-MC}(\Sigma^{\ell,m})\leq 
  p\textup{-MC}(\textup{strict-}\Sigma^{\ell,m}[2]).$
\end{lem}

Now, we are able to prove the main result of this section.

\begin{thm}\label{the:altd}For all $\ell, t\geq 1$
\[
\textup{A}[\ell,t]
=[p\textup{-MC}(\Sigma^{\ell,t-1})]^{\fpt}=[p\textup{-MC}(\Sigma^{\ell,t-1}[2])]^{\fpt} 
\]
Moreover,  we have
\begin{itemize}
\item if  $\ell$ is odd, then  
$$\textup{A}[\ell,t]=[\textsc{AWSat$_{\ell}(\Gamma_{t,2})$}]^{\fpt}
\mbox{\quad and for $t\ge 2$, \ }\textup{A}[\ell,t]=[\textsc{AWSat$_{\ell}(\Gamma_{t,1})$}]^{\fpt}
;$$
\item if  $\ell$ is even, then 
$$\textup{A}[\ell,t]=[\textsc{AWSat$_{\ell}(\Delta_{t,2})$}]^{\fpt}\mbox{\quad and for $t\ge 2$, \ }\textup{A}[\ell,t]=[\textsc{AWSat$_{\ell}(\Delta_{t,1})$}]^{\fpt}.$$
\end{itemize}
\end{thm}
Before proving this theorem, we state two consequences; the first one is the characterisation of
the A-hierarchy by means of propositional logic:
\begin{cor}\label{cor:fel}$\textup{A}[\ell]=\textup{A}[\ell,1]$, i.e.,
$\textup{A}[\ell]=[\{\textsc{AWSat}_{\ell}(\Omega_{1,d}) \mid d\geq 1\}]$.
\end{cor}
\begin{cor} For $\ell\ge 1$ and $t\ge 2$, \ $\textup{A}[\ell,t]\se \textup{A}[\ell+1,t-1]$.
\end{cor}
\proof Since $\Sigma^{\ell,t-1}\se \Sigma^{\ell+1,t-2}$, the claim follows from Theorem \ref{the:altd}.
\qed
 
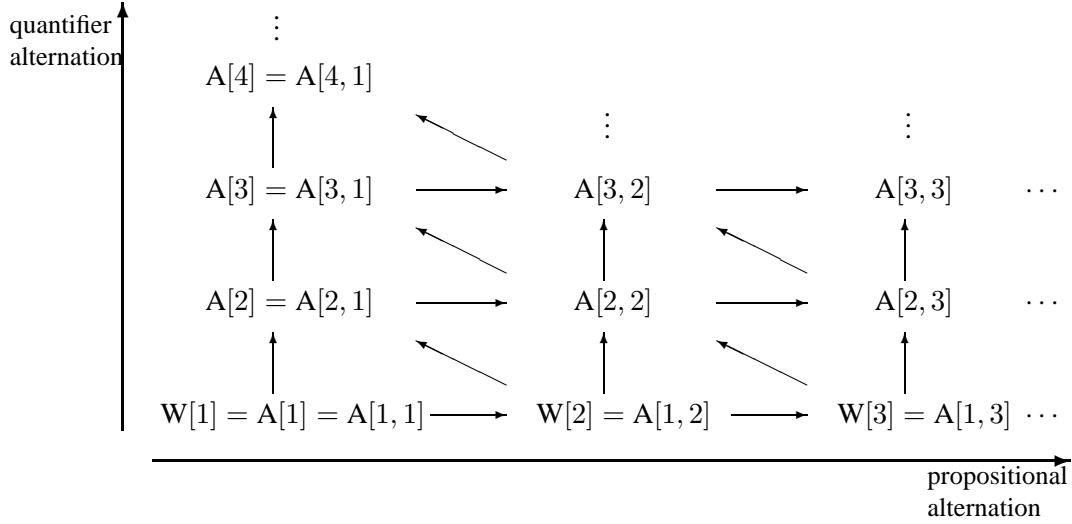
\begin{figure}
\unitlength1mm
\begin{center}
\begin{picture}(130,75)
\put(10,10){$\W{1}=\A{1}=\A{1,1}$}
\put(60,10){$\W{2}=\A{1,2}$}
\put(100,10){$\W{3}=\A{1,3}$}
\put(125,10){$\cdots$}
\put(16,25){$\A{2}=\A{2,1}$}
\put(65,25){$\A{2,2}$}
\put(105,25){$\A{2,3}$}
\put(125,25){$\cdots$}
\put(16,40){$\A{3}=\A{3,1}$}
\put(65,40){$\A{3,2}$}
\put(105,40){$\A{3,3}$}
\put(125,40){$\cdots$}
\put(16,55){$\A{4}=\A{4,1}$}
\put(69,48){$\vdots$}
\put(109,48){$\vdots$}
\put(25,61){$\vdots$}
\put(25,14){\vector(0,1){8}}
\put(25,29){\vector(0,1){8}}
\put(25,44){\vector(0,1){8}}
\put(69,14){\vector(0,1){8}}
\put(69,29){\vector(0,1){8}}
\put(109,14){\vector(0,1){8}}
\put(109,29){\vector(0,1){8}}
\put(46,11){\vector(1,0){10}}
\put(86,11){\vector(1,0){10}}
\put(44,26){\vector(1,0){12}}
\put(84,26){\vector(1,0){12}}
\put(44,41){\vector(1,0){12}}
\put(84,41){\vector(1,0){12}}
\put(56,15){\vector(-2,1){12}}
\put(96,15){\vector(-2,1){12}}
\put(56,30){\vector(-2,1){12}}
\put(96,30){\vector(-2,1){12}}
\put(56,45){\vector(-2,1){12}}
\thicklines
\put(9,5){\vector(1,0){122}}
\put(5,9){\vector(0,1){57}}
\put(112,0){\parbox{20mm}{\small propositional\\alternation}}
\put(-10,60){\parbox{20mm}{\small quan\-tifier\\alter\-nation}}
\end{picture}
\end{center}
\caption{The A-matrix of parameterized complexity classes. Arrows indicate
  containment.}\label{fig:matrix}
\end{figure}

Figure~\ref{fig:matrix} shows the matrix
and the containment relations known to hold between the classes.

Since $\textup{W}[t]=\textup{A}[1,t]$ and $\Sigma^{1,t-1}=\Sigma_{t,1}$,
Theorem \ref{the:altd} (partly) generalises Theorem \ref{the:afo} and, in fact,
its proof extends the argument given there.

\medskip\noindent 
\textit{Proof of Theorem \ref{the:altd}:} We first prove
that $\textsc{AWSat}_{\ell}(\Omega_{t,d})\le
p\textup{-MC}(\Sigma^{\ell,t-1})$.  Let
\[((\alpha,I_1,\ldots,I_\ell),(k_1,\ldots,k_\ell))\] be an instance of
$\textsc{AWSat}_{\ell}(\Omega_{t,d})$. By Lemma \ref{lem:pn1}, we may actually
assume that $\alpha\in\Delta_{t+1,d}$. Let $k:=k_1+\ldots+k_\ell$ and
$\{X_1,\ldots,X_m\}$ the set of variables of $\alpha$.

Let us first assume that $\ell$ is odd. We construct a structure $\str A$ and a
formula $\psi$ according to Lemma~\ref{lem:afo}. We expand $\mathcal A$ by
unary relation $V_1,\ldots,V_\ell$ such that $V_i^{\mathcal A}:=\{j\mid X_j\in
I_i\}$. For simplicity, we denote the resulting structure by $\mathcal A$
again. We let 
\begin{align*}
\phi&:=
\exists x_1\ldots\exists x_{k_1}\Big(
\bigwedge_{i=1}^{k_1}V_1x_i\wedge\bigwedge_{\substack{i,j=1\\i\neq
    j}}^{k_1}x_i\neq x_j\;\wedge\\
&\hspace{1cm}
\forall x_{k_1+1}\ldots\forall x_{k_1+k_2}\Big(
\big(\bigwedge_{i=k_1+1}^{k_1+k_2}V_2x_i\wedge
\bigwedge_{\substack{i,j=k_1+1\\i\neq j}}^{k_1+k_2}x_i\neq x_j\big)\to\\
&\hspace{2cm}\ldots\\
&\hspace{3cm}
\exists x_{k_1+\ldots +k_{\ell-1+1}}\ldots\exists x_{k}
\Big(
\bigwedge_{i=k_1+\ldots+k_{\ell-1}+1}^{k}V_{\ell}x_i
\wedge\bigwedge_{\substack{i,j=k_1+\ldots+k_{\ell-1}+1\\i\neq j}}^{k}x_i\neq
x_j\;\wedge\\
&\hspace{13cm}\psi\Big)\cdots\Big)\Big).
\end{align*}
It is straightforward to verify that $\mathcal A\models\phi$ if and only if,
$((\alpha,I_1,\ldots,I_\ell),(k_1,\ldots,k_\ell))$ is a `yes'-instance of
$\textsc{AWSat}_{\ell}(\Omega_{t,d})$ and that $\varphi$ is equivalent to a $\Sigma^{\ell,t-1}$-formula.

If $\ell$ is even, we assume that $\alpha\in \Gamma_{t+1,d}$ and observe that
Lemma~\ref{lem:afo} has a corresponding version for such formulas.

\bigskip By Lemma \ref{lem:nla}, it remains to get a reduction from
$p\textup{-MC}(\Sigma^{\ell,t-1}[2])$ to \textsc{AWSat$_{\ell}(\Omega_{t,d})$}
for some $d$ (and to prove the additional claims of the theorem).

First, we treat the case $t=1$ and for notational simplicity, assume $\ell=3$.
Let $\phi$ be a $ \Sigma^{3,0}[2]$-formula.
By Lemma \ref{lem:mcr}(2), we may assume that $\varphi$ is a simple $\Sigma_{3}$-sentence,
\[
\phi:=\exists x_1\ldots \exists x_h\forall y_1\ldots\forall  y_k\exists z_1\ldots \exists z_m(\lambda_{1}\land\ldots \land\lambda_{s})
\]
with literals $\lambda_{i}$ and  $\mathcal A$ a structure  in the corresponding vocabulary.

We first construct a propositional formula $\alpha'\in \Omega_{1,d}$ for some $d$. For the partition of its propositional variables into the three sets
\[
I_1:=\{X_{i,a} \mid i=1,\ldots, h,\, a\in A\}, \quad  I_2:=\{Y_{i,a} \mid i=1,\ldots, k,\, a\in A\},  \]
and
\[  I_3:=\{Z_{i,a} \mid i=1,\ldots, m,\, a\in A\},
\] and for the natural numbers   $h,k,m$, we will see that
\begin{eqnarray}\label{eqn:a.3re}\str{A}\models\phi&\iff&((\alpha',I_1, I_2, I_3),(h,k,m))\in p\textsc{-AWSat}_{\ell}(\Omega_{1,d}).
\end{eqnarray}

Clearly, the intended meaning of $X_{i,a}$ is ``$x_i$ gets the value $a$'' and similarly for the other variables. 

The formula $\alpha' $ has the form $(\bigwedge\ldots  \lor \bigvee ...)$. The ``big'' conjunction takes care of existentially quantified variables: it contains as conjuncts $(\neg X_{i,a}\lor \neg X_{i,b})$ for $i=1,\ldots, h$,  $ a,b\in A,\ a\not= b$ and  $(\neg Z_{i,a}\lor \neg Z_{i,b})$ for $i=1,\ldots, m$,  $ a,b\in A,\ a\not= b$. The ``big'' disjunction takes care of universally quantified variables; in fact, it {\em only} contains  as disjuncts $(Y_{i,a}\land Y_{i,b})$  for $i=1,\ldots, k$,  $ a,b\in A,\ a\not= b$.  So far, it should be clear that any satisfying assignment of $\alpha'$ of ``weight $h,k,m$'' sets 
\begin{itemize}
\item for every $i$ exactly one variable $X_{i,a}$ to \textsc{True} and similarly for the $ Z_{i,a}$
\end{itemize}
or 
\begin{itemize}
\item  it sets $Y_{i,a}, Y_{i,b}$ to \textsc{True} for some $i$ and some $a,b\in A,\ a\not= b $.
\end{itemize}
Finally, we take care of the quantifier-free part of  $\phi$ by adding to the big conjunction for every $\lambda_{i}$, say $\lambda_{i}(x_3,y_2)$ (recall that the arity of the vocabulary is $\le 2$), and every $(a,b)\in A$ with $\str{A}\not\models \lambda_{i}(a,b)$ as conjunct the formula $( \neg X_{3,a}\lor  \neg Y_{2,b}) $. We leave the verification of \eqref{eqn:a.3re} to the reader.

Now, we show how to get rid of the big disjunction in $\alpha'$, thus proving
the additional claim 
\[
p\textup{-MC}(\Sigma^{3,0}[2])\le
\textsc{AWSat}_{3}(\Gamma_{1,2}).
\] 
Besides the propositional variables of
$\alpha'$, the formula $\alpha$ we aim at has additional propositional
variables, namely the variables
\[
C,Y_1,\ldots, Y_k,Z_1,\ldots, Z_m.
\]
The partition of the  variables of $\alpha$  consists of  three sets, namely of $I_1$ and $I_2$ as above, i.e.,
\[
I_1:=\{X_{i,a} \mid i=1,\ldots, h,\, a\in A\}, \quad  I_2:=\{Y_{i,a} \mid i=1,\ldots, k,\, a\in A\},\]
and of $J_3$ that  contains the variables  of $I_3$ and the new variables, i.e.,
\[  J_3:=\{Z_{i,a} \mid i=1,\ldots, m,\, a\in A\}\cup\{C,Y_1,\ldots, Y_k,Z_1,\ldots, Z_m \}.
\] The ``parameters'' are $h,k,m+1$. In fact we will have
\begin{equation}\label{equ:obd}
\begin{array}{c@{}l}
&((\alpha',I_1, I_2, I_3),(h,k,m))\in
\textsc{AWSat}_{3}(\Omega_{1,d})\\
\iff&((\alpha,I_1, I_2, J_3),(h,k,m+1))\in
\textsc{AWSat}_{3}(\Gamma_{1,2}).
\end{array}
\end{equation}
To understand the construction of $\alpha$ better, we briefly explain the meaning or role of the new propositional variables: $C$ essentially signalizes that the big conjunction in $\alpha'$ is satisfied, $Y_i$ that no variable $Y_{i,a}$ with $a\in A$ has been chosen; finally, in case the big disjunction in $\alpha'$ is satisfied, then $Z_1,\ldots, Z_m$, but no $Z_{i,a}$, will be set to \textsc{true}.

Let $\alpha$ be obtained from $\alpha'$ by 
\begin{itemize}
\item eliminating the big disjunction; 
\item adding to the big conjunction the formulas (the indices always range over all possible values)
\begin{enumerate}
\item $\neg C\vee\neg Y_i$, 
\item $\neg C\vee\neg Z_i$
\item $\neg Z_i\vee\neg Z_{j,a}$
\item $\neg Y_i\vee\neg Z_{j,a}$
\item $\neg Y_i\vee\neg Y_{i,a}$
\item $\neg Y_i\vee\neg Y_{j}$ for $i\not= j$.
\end{enumerate}
\end{itemize}
Then, $\alpha$ is  in $\Gamma_{1,2}$. We verify (\ref{equ:obd}).
Assume first that $(\alpha',V_1, V_2, V_3,h,k,m)\in
\textsc{AWSat}_{3}(\textsc{Prop})$. To verify the right hand side of 
(\ref{equ:obd}), we choose $S_1\se I_1$ as it is done when verifying the left side. Now let $S_2$ be any size $k$ subset of $ I_2$; if  $S_2$ does not satisfy $\bigvee_{\ldots }(Y_{i,a}\wedge Y_{i,b})$,
then we select $S'_3\se I_3$ as when verifying for $S_1,S_2$ the left hand side. Then, we can set $S_3:=S'_3\cup \{C\}$ and verify that $S_1\cup S_2\cup S_3$ satisfies $\alpha$. If $S_2$  satisfies $\bigvee_{\ldots }(Y_{i,a}\wedge Y_{i,b})$,  then there is some $i_0$ such that $Y_{i_0,a}\notin S_2$ for all $a\in A$. We set $S_3:=\{Y_{i_0},Z_1,\ldots, Z_m \}$ and again verify that $S:=S_1\cup S_2\cup S_3$ satisfies $\alpha$. Clearly, $S$ satisfies all clauses (1)--(6). And, it also satisfies all old conjuncts, since they are not of the form $(\neg Y_{i,a}\vee\neg Y_{i,b})$.

Conversely, assume that the right hand side of (\ref{equ:obd}) holds. For $\alpha'$ we choose $S_1$ as it is done for $\alpha$ when verifying the right hand side. Let $S_2$ be any size $k$ subset of $ I_2$; if  $S_2$  satisfies $\bigvee_{\ldots }(Y_{i,a}\wedge Y_{i,b})$ we are done. Otherwise, we choose for  $S_1,S_2$ a size $m+1$ subset $S_3$ of $J_3$ such that $S_1\cup S_2\cup S_3$ satisfies $\alpha$. By the formulas  (5), $S_3$ does not contain any $Y_i$. By the clauses  (2) , $S_3$ at most contains $m$ variables from $\{C,Z_1,\ldots, Z_m  \}$. Therefore,  for some $j$ there is at least one $a\in A$ such that $Z_{j,a}\in S_3$. But then, by the clauses (3), the set $S_3$ contains no $Z_i$. Thus, $S_3$ contains $C$ and for every $j$ exactly one $Z_{j,a}$ (recall that the big conjunction in $\alpha'$ and hence, the one in $\alpha$, contains the conjuncts $(\neg Z_{i,a}\vee\neg Z_{i,b}) $ for $i=1,\ldots, m$ and $a,b\in A$ with $a\not= b$). Therefore,  setting $S_3':=S_3\cap I_3$, we have  $S_1\cup S_2\cup S_3'$ satisfies $\alpha'$.
\medskip

Now, let us assume that $t\geq 2$ and, say, $\ell$ is odd. We aim at a reduction to $\textsc{AWSat}_{\ell}(\Gamma_{t,1})$. The formula $\phi$ has the form
\[
\exists \bar x_1\forall \bar x_2\ldots \exists\bar x_{\ell}\forall x_{\ell+1}\ldots Q_{\ell+(t-1)}x_{\ell+(t-1)}\psi,
\]
i.e., the first ``short'' quantifier block (consisting of a single quantifier) is universal. Moreover, we can assume that $\varphi$ is strict, that is, that every atomic subformula contains at most one variable of the unrestricted block. 
The unrestricted blocks are treated in the propositional formula  as above and the short blocks and the quantifier-free part as in the proof of Proposition \ref{pro:altd1}.
In particular, to the big conjunction of the propositional formula $\alpha'=(\bigwedge\ldots  \lor \bigvee ...)$ constructed  for $t=1$, we add conjuncts corresponding to the quantifier $\forall x_{\ell +1}$. Below  this big conjunction there is a layer of big disjunctions. (In case $t\ge 3$ this layer can also be used to eliminate the big disjunction of $$\alpha'=(\bigwedge\ldots  \lor \bigvee_{i=1,\ldots, k; a,b\in A, a\neq b}(Y_{i,a}\wedge Y_{i,b})),$$
which is treated as a $\Delta_{2,1}$-formula.) 
We argue as above to get rid of the big disjunction of $\alpha'$.

 Altogether, we obtain a reduction to
 $\textsc{AWSat}_{\ell}(\Gamma_{t,1})$. Similarly, one argues in case $\ell$
 is even: Then the first ``short'' quantifier block is existential, and therefore one  obtains a reduction to\\$\textsc{AWSat}_{\ell}(\Delta_{t,1})$.
\qed

Arguing as in the derivation of Corollary \ref{cor:del}, one obtains

\begin{rem}For $t\ge 2$ and $d\ge 1$,
\begin{itemize}
\item if $\ell$ is odd, then $\textsc{AWSat$_{\ell}(\Delta_{t,d})$}\in
  \textup{A}[\ell,t-1]; $
\item if $\ell$ is even, then, $\textsc{AWSat$_{\ell}(\Gamma_{t,d})$}\in
  \textup{A}[\ell,t-1] .$
\end{itemize}
\end{rem} 
\begin{rem}
  As for the W-hierarchy one can obtain improvements restricting the
  propositional formulas to monotone or antimonotone ones. We leave the details to the reader.
\end{rem}

\begin{rem}
  For the A-hierarchy there are two more or less natural ways to define a
  starred version $\textup{A}^*[1],\textup{A}^*[2],\ldots $. From the point of
  view of first-order logic, we introduce the classes of formulas $\Sigma^*_t$
  by induction
  \[
  \begin{array}{rcl}
    \Sigma^*_0&:=&\parbox[t]{8cm}{the set of quantifier-free formulas}\\
    \Sigma^*_{t+1}&:=&\parbox[t]{8cm}{formulas of the form 
      $\exists y_1\ldots \exists y_u \psi$, where $\psi$ is a 
      Boolean combination of formulas in $ \Sigma^*_t$,}
  \end{array}
  \]
  and set 
  \[
  \textup{A}^*[t]:=[p\textup{-MC}(\Sigma^*_t)]^{\fpt}.
  \]
  But since every formula in $\Sigma^*_t$ is logically equivalent to a formula
  in $\Sigma_t$, we immediately get $\textup{A}^*[t]=\textup{A}[t]$.
  
  From the point of view of propositional logic we imitate the definition of
  W$^*$ in the alternating context: For a set $\Gamma$ of propositional
  formulas let \npprob{ \textsc{AWSat$^*_{\ell}(\Gamma)$}}{$\alpha\in\Gamma$,
    $k\in \mathbb N$ such that the depth of $\alpha$ is at most $k$, and a
    partition $I_1\, \dot{\cup}\, \ldots \, \dot{\cup}\, I_{\ell}$ of the
    propositional variables of $\alpha$}{$k_1,\ldots ,k_{\ell}\in\mathbb N$
    with $k=k_1+\ldots +k_{\ell}$.}{Decide if there is a size $k_1$ subset
    $S_1$ of $I_1$ such that for every size $k_2$ subset $S_2$ of $I_2$ there
    exists \ldots such that the assignment $S_1\cup\ldots \cup S_{\ell}$
    satisfies $\alpha$} And set
\[
\textup{A}^*[t]:=[\{\textsc{AWSat}^*_{t}(\Omega_{1,d})\mid d\geq 1\}]^{\fpt}.
\]
Clearly,   $\textsc{AWSat}_{t}(\Omega_{1,d})\leq \textsc{AWSat}^*_{t}(\Omega_{1,d})$.  On the other hand, essentially the proof of  Proposition \ref{the:altd} shows that $\textsc{AWSat}^*_{t}(\Omega^*_{1,d})\leq p\textup{-MC}(\Sigma^*_t) $, so that again we obtain $\textup{A}^*[t]=\textup{A}[t]$.
\end{rem}

\subsection{The AW-hierarchy}\label{sub:aw} 
Downey and Fellows \cite{dowfel99} introduced the AW-hierarchy and
showed its collapse. Again this result can easily be derived (and
slightly be improved) with the techniques developed in this paper.

To define this hierarchy, for a set $\Gamma$ of propositional formulas, we
introduce the {\em alternating weighted satisfiability problem}
\textsc{AWSat$(\Gamma)$} (in contrast to \textsc{AWSat$_{\ell}(\Gamma)$}
defined in the preceding section we have no restriction on the number of
alternations): \npprob{ \textsc{AWSat$(\Gamma)$}}{$\alpha\in\Gamma$, $\ell\ge
  1$, and a partition $I_1\, \dot{\cup}\, \ldots \, \dot{\cup}\, I_{\ell}$ of
  the propositional variables of $\alpha$}{$k_1,\ldots ,k_{\ell}\in\mathbb
  N$.}{Decide if there is a size $k_1$ subset $S_1$ of $I_1$ such that for
  every size $k_2$ subset $S_2$ of $I_2$ there exists \ldots such that the
  truth assignment $S_1\cup\ldots \cup S_{\ell}$  satisfies $\alpha$} Hence, given the input
$(\alpha,\ell,I_1\, \dot{\cup}\, \ldots \, \dot{\cup}\, I_{\ell})$ and the
parameter $(k_1,\ldots ,k_{\ell})$ we have the equivalence
\begin{equation}\label{eqn:awl}
\begin{array}{c@{}l}
&((\alpha,\ell,I_1, \ldots , I_{\ell}),(k_1,\ldots ,k_{\ell}))\in \textsc{AWSat}(\Gamma)\\
\iff&((\alpha,I_1, \ldots, I_{\ell}),(k_1,\ldots ,k_{\ell}))\in \textsc{AWSat}_\ell(\Gamma)
\end{array}
\end{equation}
(note that on the left side of the equivalence the number  $\ell$ is part of the input and is not fixed in advance).
\begin{defi}For $t\ge 1$, $\textup{AW}[t]:=[\{\textsc{AWSat}(\Gamma_{t,d})\mid d\ge 1\}]^{\fpt}$.
\end{defi}In a very informal way the core of the proof of the following theorem can be described in the following form:
\begin{eqnarray*}\textup{AW}[t]&=&[\{\textup{``$\bigcup_{\ell\ge 1}\textsc{AWSat}_\ell(\Gamma_{t,d})$''}\mid d\ge 1\}]^{\fpt} \mbox{\qquad by (\ref{eqn:awl})}\\
&=&[ p\textup{-MC}(\bigcup_{\ell\ge 1}\Sigma^{\ell,t-1})]^{\fpt} \mbox{\qquad by Theorem \ref{the:altd}.}
\end{eqnarray*}
Since $\bigcup_{\ell\ge 1}\Sigma^{\ell,t-1}=\bigcup_{\ell\ge 1}\Sigma^{\ell,0}=\textup{FO}$, we get $\textup{AW}[1]=\textup{AW}[t]=[p\textup{-MC(FO)}]^\fpt$, which essentially is the statement of the following theorem.

\begin{thm}For $t\ge 1$,
\[
\textup{AW}[1]=\textup{AW}[t]=[\textsc{AWSat}(\Gamma_{1,2})]^{\fpt}=[p\textup{-MC(FO)}]^{\fpt}=[p\textup{-MC(FO[2])}]^{\fpt}.
\]
\end{thm}

\proof Clearly, $\textup{AW}[1]\se\textup{AW}[t]$. Consider an instance of
$\textsc{AWSat}(\Gamma_{t,d})$ consisting of the input \[(\alpha,\ell,I_1,
\ldots , I_{\ell})\] and the parameter $(k_1,\ldots ,k_{\ell})$. In the proof
of Theorem \ref{the:altd} we saw how to proceed in order to obtain a structure
\str{A} and a formula $\phi\in \Sigma^{\ell,t-1}$ such that
\begin{eqnarray*}((\alpha,\ell,I_1, \ldots , I_{\ell}),(k_1,\ldots ,k_{\ell}))\in\textsc{AWSat}(\Gamma_{t,d})&\iff&\str{A}\models \phi. 
\end{eqnarray*}
Clearly, this procedure is uniform in $\ell$ and  an \fpt-reduction from $\textsc{AWSat}(\Gamma_{t,d})$ to  $p\textup{-MC(FO)}$.
By part (3) of the First-Order Normalisation Lemma, we know that $p\textup{-MC(FO)}\le p\textup{-MC(FO[2])}$.
 Finally, let  $\str{A}$ be a structure and $\phi\in \textup{FO[2]}$a formula, say $\phi\in\Sigma_{\ell}=\Sigma^{\ell,0}$. We may  assume that $\ell$ is odd. Then the proof of Theorem \ref{the:altd} shows how to obtain a formula $\alpha\in\Gamma_{1,2}$, a partition $I_1\, \dot{\cup}\, \ldots \,
  \dot{\cup}\, I_{\ell}$ of its variables,  and $k_1,\ldots, k_{\ell}$ such that
\begin{eqnarray*}
\str{A}\models \phi&\iff&((\alpha,I_1, \ldots , I_{\ell}),(k_1,\ldots ,k_{\ell}))\in \textsc{AWSat}_\ell(\Gamma_{1,2}),
\end{eqnarray*}
i.e., such that
 \begin{eqnarray*}
\str{A}\models \phi&\iff&((\alpha,\ell,I_1, \ldots , I_{\ell}),(k_1,\ldots ,k_{\ell}))\in \textsc{AWSat}(\Gamma_{1,2}).
\end{eqnarray*}
Hence,  we have an \fpt-reduction from $p\textup{-MC(FO[2])}$ to $\textsc{AWSat}(\Gamma_{1,2})$. \qed

\section{Conclusions}
We hope to have demonstrated that the correspondence between propositional and
first-order logic, or more precisely, weighted satisfiability and
model-checking problems, is very fruitful. We see this correspondence
at the core of structural parameterized complexity theory. Once it is
established, many other results follow quite easily.

Several problems remain open, the most important being the question of whether
the W-hierarchy and the $\text{W}^*$-hierarchy coincide. Even though our
results clarify what is known, we have failed to make any definite progress on
this problem.

Another nagging open question is whether the First-Order Normalisation Lemma
can be extended to vocabularies with function symbols. A positive answer would
greatly simplify the machine characterisation of the classes of the
W-hierarchy given in \cite{cheflu03}.


\begin{thebibliography}{10}

\bibitem{abrdowfel95}
K.A. Abrahamson, R.G. Downey, and M.R. Fellows.
\newblock Fixed-parameter tractability and completeness {IV}: {O}n completeness
  for {W[P]} and {PSPACE} analogs.
\newblock {\em Annals of pure and applied logic}, 73:235--276, 1995.

\bibitem{cheflu03}
Y.~Chen and J.~Flum.
\newblock Machine characterizations of the classes of the {W}-hierarchy.
\newblock In M.~Baaz and J.~Makowsky, editors, {\em Proceedings of the 17th
  International Workshop on Computer Science Logic}, volume 2803 of {\em
  Lecture Notes in Computer Science}, pages 114--127. Springer-Verlag, 2003.

\bibitem{dowfel95b}
R.G. Downey and M.R. Fellows.
\newblock Fixed-parameter tractability and completeness {I}: Basic results.
\newblock {\em {SIAM} Journal on Computing}, 24:873--921, 1995.

\bibitem{dowfel95}
R.G. Downey and M.R. Fellows.
\newblock Fixed-parameter tractability and completeness {II}: On completeness
  for ${W}[1]$.
\newblock {\em Theoretical Computer Science}, 141:109--131, 1995.

\bibitem{dowfel98}
R.G. Downey and M.R. Fellows.
\newblock Threshold dominating sets and an improved characterization of
  {$W[2]$}.
\newblock {\em Theoretical Computer Science}, 209:123--140, 1998.

\bibitem{dowfel99}
R.G. Downey and M.R. Fellows.
\newblock {\em Parameterized Complexity}.
\newblock Springer-Verlag, 1999.

\bibitem{dowfelreg98}
R.G. Downey, M.R. Fellows, and K.~Regan.
\newblock Descriptive complexity and the ${W}$-hierarchy.
\newblock In P.~Beame and S.~Buss, editors, {\em Proof Complexity and Feasible
  Arithmetic}, volume~39 of {\em AMS-DIMACS Volume Series}, pages 119--134.
  AMS, 1998.

\bibitem{dowfeltay96}
R.G. Downey, M.R. Fellows, and U.~Taylor.
\newblock The parameterized complexity of relational database queries and an
  improved characterization of ${W}[1]$.
\newblock In D.S. Bridges, C.~Calude, P.~Gibbons, S.~Reeves, and I.H. Witten,
  editors, {\em Combinatorics, Complexity, and Logic -- Proceedings of DMTCS
  '96}, pages 194--213. Springer-Verlag, 1996.

\bibitem{fag74}
R.~Fagin.
\newblock Generalized first--order spectra and polynomial--time recognizable
  sets.
\newblock In R.~M. Karp, editor, {\em Complexity of Computation, SIAM-AMS
  Proceedings, Vol. 7}, pages 43--73, 1974.

\bibitem{flufrigro02}
J.~Flum, M.~Frick, and M.~Grohe.
\newblock Query evaluation via tree-decompositions.
\newblock {\em Journal of the ACM}, 49(6):716--752, 2002.

\bibitem{flugro01}
J.~Flum and M.~Grohe.
\newblock Fixed-parameter tractability, definability, and model checking.
\newblock {\em {SIAM} Journal on Computing}, 31(1):113--145, 2001.

\bibitem{frigro03}
M.~Frick and M.~Grohe.
\newblock The complexity of first-order and monadic second-order logic
  revisited.
\newblock \emph{Annals of Pure and
  Applied Logic}, 130:3--31, 2004.

\bibitem{gro01c}
M.~Grohe.
\newblock The parameterized complexity of database queries.
\newblock In {\em Proceedings of the 20th ACM Symposium on Principles of
  Database Systems}, pages 82--92, 2001.

\bibitem{licpnu85}
O.~Lichtenstein and A.~Pnueli.
\newblock Finite state concurrent programs satisfy their linear specification.
\newblock In {\em Proceedings of the Twelfth ACM Symposium on the Principles of
  Programming Languages}, pages 97--107, 1985.

\bibitem{siscla84}
A.P. Sistla and E.M. Clarke.
\newblock The complexity of propositional linear temporal logic.
\newblock {\em Journal of the ACM}, 32(3):733--749, 1985.

\bibitem{sto74}
L.J. Stockmeyer.
\newblock {\em The Complexity of Decision Problems in Automata Theory}.
\newblock PhD thesis, Department of Electrical Engineering, MIT, 1974.

\bibitem{var82}
M.Y. Vardi.
\newblock The complexity of relational query languages.
\newblock In {\em Proceedings of the 14th ACM Symposium on Theory of
  Computing}, pages 137--146, 1982.

\end{thebibliography}
\end{document}